\documentclass[%
 reprint,
 amsmath,amssymb,
 aps,
 onecolumn,  
longbibliography
]{revtex4-2}
\usepackage{graphicx}
\usepackage{dcolumn}
\usepackage{scalerel}
\usepackage{stackengine,wasysym}
\usepackage{dsfont}
\usepackage{bm}
\usepackage{braket}
\usepackage{here}
\usepackage[toc,page]{appendix}

\begin{document}

\preprint{APS/123-QED}

\title{Blueprint for
quantum computing using electrons on helium}
\author{Erika Kawakami$^{1,2,*}$, Jiabao Chen$^{3}$, Mónica Benito$^{4}$, Denis Konstantinov$^{5}$}
\affiliation{$^1$RIKEN Center for Quantum Computing, RIKEN, Wako, 351-0198, Japan}
\affiliation{$^2$Cluster for Pioneering Research, RIKEN, Wako, 351-0198, Japan}
\email[E-mail: ]{e2006k@gmail.com}
\affiliation{$^3$QunaSys Inc., Bunkyo, Tokyo 113-0001, Japan}
\affiliation{$^4$Institute of Quantum Technologies, German Aerospace Center (DLR), Wilhelm-Runge-Straße 10, 89081, Ulm, Germany}
\affiliation{$^5$Quantum Dynamics Unit, Okinawa Institute of Science and Technology (OIST) Graduate University, Okinawa 904-0495, Japan}





\date{\today}

\begin{abstract}
We present a blueprint for building a fault-tolerant quantum computer using the spin states of electrons on the surface of liquid helium. We propose to use ferromagnetic micropillars to trap single electrons on top of them and to generate a local magnetic field gradient. Introducing a local magnetic field gradient hybridizes  charge and spin degrees of freedom, which allows us to benefit from both the long coherence time of the spin state and the long-range Coulomb interaction that affects the charge state. We present concrete schemes to realize single- and two-qubit gates and quantum-non-demolition
read-out. In our framework, the hybridization of charge and spin degrees of freedom is large enough to perform fast qubit gates and small enough not to degrade the coherence time of the spin state significantly, which leads to the realization of high-fidelity qubit gates. 
\end{abstract}
\pacs{Valid PACS appear here}
\maketitle

\section{Introduction}

Fault-tolerant quantum computing requires a high enough number of interacting qubits placed in two dimensions~\cite{Fowler2012}. Preparing such a system while realizing high-fidelity qubit operations is indispensable. The interest in using electrons in vacuum as qubits has been growing recently. As opposed to using electrons trapped in semiconductor structures, such systems are free of defects or impurities and we can expect long coherence times of their quantum states, which tends to lead to  high-fidelity qubit operations. Historically, a single electron was first trapped in vacuum using a Penning trap~\cite{Wineland1973}. Recent efforts on trapping electrons in Paul traps in terms of using them as qubits are reported~\cite{Matthiesen2021-zk,Yu2022-ma}. A remarkable experimental  milestone has been recently achieved using electrons on the surface of solid neon by demonstrating the single-qubit gate operations with 99.95\% fidelity~\cite{Zhou2022-nk,Zhou2022-de}. Electrons on the surface of liquid helium is another physical system representing electrons in vacuum. This two-dimensional electron system is known for having the highest measured mobility~\cite{Shirahama1995-qo} thanks to the clean interface between liquid helium and vacuum, which suggests the potential for forming a substantial number of uniform qubits. The hydrodynamic instability of liquid can be suppressed for the liquid helium confined in the micro-fabricated devices filled by capillary action~\cite{Marty1986-om,Rousseau2007,Ikegami2009-xf,Zou2022-fu,Rees2011-eq,Rees2012-zi}. Besides the theoretical proposals of using electrons on helium as qubits in early days~\cite{Platzman1999,Dykman2003,Lyon2006,Schuster2010}, experimental studies on electrons on helium with the aim of using them as qubits are also reported such as trapping a few number of electrons \cite{Papageorgiou2005}, shuttling of the electrons~\cite{Bradbury2011-ec}, coupling of electrons to microwave (MW) photons via superconducting resonator~\cite{Koolstra2019-mq}, and coupling of electrons to surface acoustic waves~\cite{Byeon2021-ht}. However, no qubit operations on those electrons have  yet been experimentally demonstrated.

The initial proposal to realize qubits using electrons on helium focused on the quantized bound states of the electron orbital motion perpendicular to the liquid surface~\cite{Platzman1999,Dykman2003}. 
These states of 1D motion in the $z$ direction are formed due to the interaction of an electron with its image charge in liquid helium and have a hydrogen-like energy spectrum \cite{Monarkha2004Two-DimensionalSystems}. The corresponding eigenfunctions are similar to the radial part of the eigenfunctions in the hydrogen atom, therefore these eigenstates are usually referred to as the Rydberg states~\footnote{In atomic physics, a Rydberg state refers to a state of an atom or molecule in which one of the electrons has a high principal quantum number orbital. However, in the field of electrons on helium, the ground state is also called a Rydberg state.}, with a typical Rydberg constant on the order of 1~meV. The Rydberg-ground state and the Rydberg-1st-excited state are utilized as qubit states. The advantage of using the Rydberg state as the qubit state is a long-range interaction between distant qubits. A two-qubit gate can be completed via the coupling of qubits by the electric dipole-dipole interaction which arises from the Coulomb interaction between electrons. It utilizes the fact that since an electron's vertical position depends on the Rydberg state, so does the strength of the electric dipole-dipole interaction. This is a unique feature of electrons on cryogenic substrates such as liquid helium, solid neon, and solid hydrogen~\cite{Monarkha2004Two-DimensionalSystems,Zhou2022-nk}.
Compared to the electrons on cryogenic substrates, the Coulomb interaction is reduced in semiconductors since the relative electric permittivity in Si and GaAs is typically $\epsilon_r~\approx 12$. Additionally, the electrons in semiconductors are more tightly confined normal to the interface. Consequently, the electrons' positions, and thus the strength of the electric dipole-dipole interaction, depend little on the vertical quantum state in semiconductors.  As shown below, the electric dipole-dipole interaction energy stays as large as 4$J/h\approx 140$~MHz for electrons on helium even if two qubits are separated by a distance as far as $d=$0.88 $\mu$m. This fact indicates that qubits can interact with each other without having any additional structures such as a floating gate \cite{Trifunovic2012} or a superconducting resonator \cite{Samkharadze2018,Koolstra2019-mq,Zhou2022-nk}, ergo  reducing the complexity of the qubit architecture.  Ref.~\cite{Platzman1999,Dykman2003} theoretically showed that a metallic pillar can trap a single electron. Cutting-edge nanofabrication techniques allow us to wire pillars every $0.88~\mu$m~\cite{Veldhorst2017}. However, in order to cancel out the unintended state-evolution caused by this always-on electric dipole-dipole interaction, one should continuously apply a decoupling sequence~\cite{Ladd2005}, which would add complications when realizing quantum gates. Another disadvantage of using the Rydberg state is its relatively short relaxation time ($T_\mathrm{1,Ry} \sim 1  \mu$s)~\cite{Monarkha2010DecayRipplons,Kawakami2021}.

The spin states of the electrons on liquid helium are expected to have an extremely long coherence time $>$100~s, and therefore  were proposed to be used as qubit states~\cite{Lyon2006}.  The long coherence time is thanks to the small magnetic dipole moment which hardly couples to the environment and thus is affected by the environmental noise little. However, the small magnetic dipole moment makes it difficult to read out the spin state. Furthermore, 
the magnetic dipole-dipole interaction between distant electrons is $<1$~Hz for $d=0.88~
\mu$m, which makes the two-qubit gate speed utilizing this interaction slow.  Later,  Schuster et al. proposed  to couple the MW photons and the orbital motion parallel to the liquid surface (hereafter, we refer to the parallel orbital motion as the orbital state, while the perpendicular orbital motion is called the Rydberg state) via a superconducting resonator~\cite{Schuster2010}. There, an external magnetic field is applied parallel to the liquid surface and the orbital and spin degrees of freedom can be coupled due to a local magnetic field gradient created by a current running through a wire underneath the electron. In this way, the spin qubit can be read out and different spin qubits can be coupled via the superconducting resonator. Alternatively, the spin states of adjacent electrons can interact with each other via the coupling of the spin state of each electron to a normal mode of the collective in-plane vibration of the  electrons arising due to the Coulomb interaction~\cite{Dykman2023-hx,Zhang2012}.

Here, we propose a hybrid qubit comprising the charge state and the spin state.  A magnetic field gradient created by a nanofabricated ferromagnet pillar introduces the interaction between  charge and spin degrees of freedom. The charge degrees of freedom refers to the in-plane orbital motion (orbital state) and the vertical orbital motion (Rydberg state). The long coherence time of the spin state is essential to realize high-fidelity qubit operations, while the long-range Coulomb interaction affecting the Rydberg states allows us to place electrons at a moderate distance while keeping a considerable interaction between them, which is a basic requirement for realizing a high number of qubits in a two-dimensional array. Single-qubit manipulation is realized by hybridizing the orbital state and the spin state.

\section{Physical realization of the hybrid charge-Spin qubit}\label{sec:physical_realization}

\begin{figure}[H]
\centering
\includegraphics[width=0.7\textwidth]{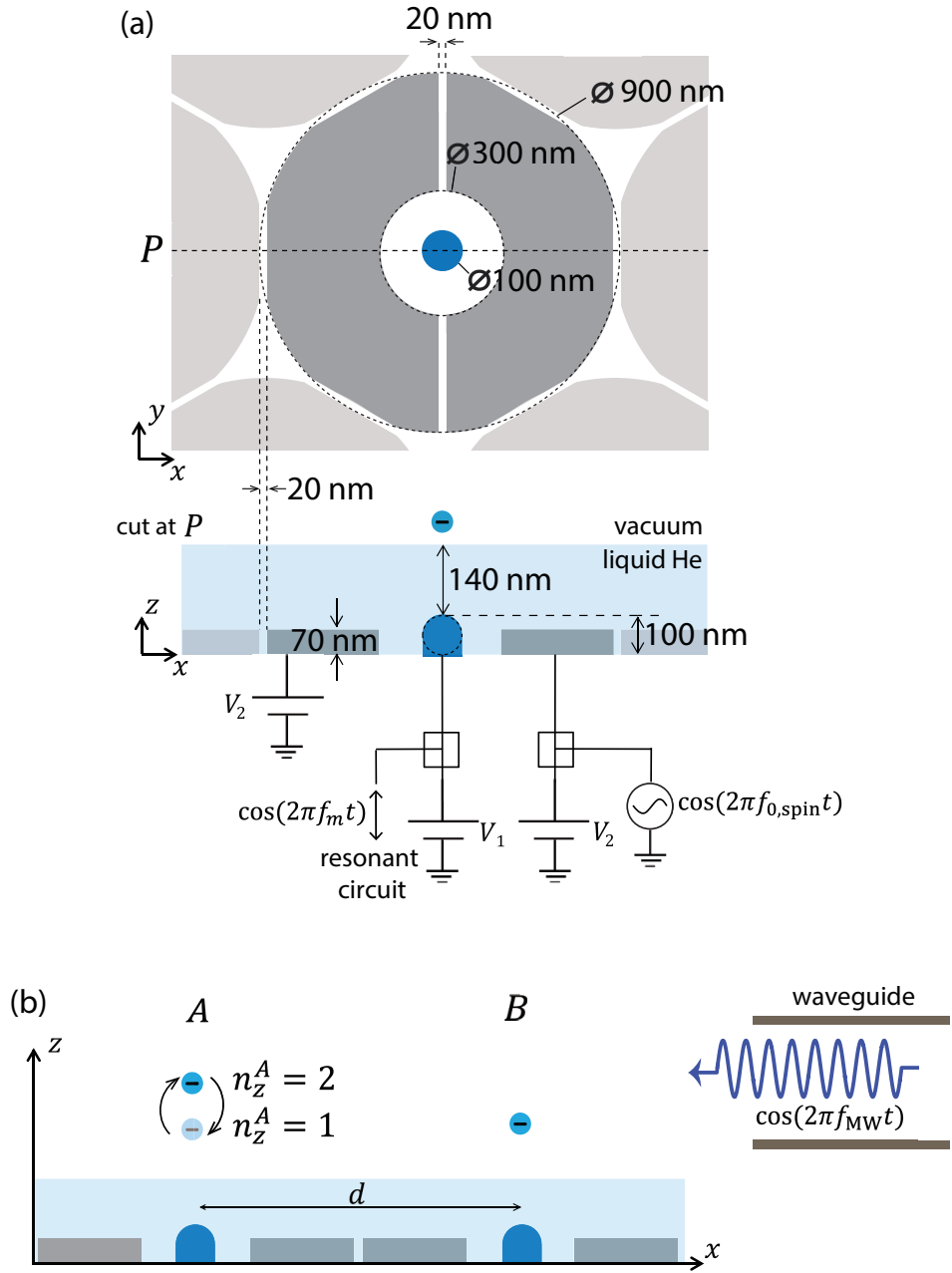}
\caption{(a) A ferromagnetic pillar (center electrode) is shown in blue and the segmented toroidal electrodes (outer electrodes)  are shown in gray. A single electron is trapped on top of the center electrode. The outer electrodes for the neighboring electrons are shown in light gray. The gap between the outer electrodes is 20~nm. The thickness of liquid $^4$He (light blue) above the center electrode is set to 140~nm. The center electrode forms a 100~nm diameter pillar, the tip of which is rounded. $V_1$ and $V_2$ are the voltages applied to the center electrode and the two outer electrodes, respectively.  An AC voltage $\propto \cos( 2 \pi f_\mathrm{0,spin} t)$ is sent to one of the outer electrodes to realize EDSR. By connecting an inductance to the center electrode, an LC resonant circuit can be formed. The resonant circuit is used to measure the spin state by sending an AC voltage $\propto \cos(2 \pi f_m t)$ to it.  (b) Two neighboring electrons A and B are separated by $d=0.88~\mu$m and are used as qubit A and qubit B, respectively. In the case shown here, electron A's Rydberg transition is on resonance with the MW applied through a waveguide. Electron A goes back and forth between the Rydberg-ground state ($n_z^A=1$) and the Rydberg-1st-excited state ($n_z^A=2$)  (see Sec.~\ref{sec:two-qubit} and Sec.~\ref{sec:detection} for details).  \label{fig:electrode_dimension_slanting_magnetic_field}}
\end{figure}

We propose to fabricate a pillar (center electrode) and segmented toroidal electrodes (outer electrodes) to trap a single electron on top of the pillar (Fig.~\ref{fig:electrode_dimension_slanting_magnetic_field}(a)).  For electrons on helium, it is well known that electrons  freeze into a Wigner crystal when the Coulomb energy $e^2/d \sim 10$~K (with $d=0.88\mu$m) exceeds the kinetic energy $k_B T $~\cite{Grimes1979-rv}. By patterning pillars into a triangular lattice, we can avoid incommensurability between the lattice formed by the pillars and the Wigner lattice formed by the Coulomb interaction. This allows us to perform topological quantum error correction for qubits in a triangular lattice ~\cite{Krantz2016-ae}. The adjacent pillars are separated by $d=0.88~\mu$m (Fig.~\ref{fig:electrode_dimension_slanting_magnetic_field}(b)) and so are the electrons trapped by them, which allows us to prepare $>10^7$  qubits in the area of 1~cm$^2$. The intrinsic interaction between  charge and spin degrees of freedom is negligibly small~\cite{Lyon2006}. Here, we introduce an artificial one via a local magnetic field gradient which is created by making a pillar of a ferromagnetic material (cobalt)~\cite{Tokura2006,Pioro-Ladriere2008} (Fig.~\ref{fig:electrode_dimension_slanting_magnetic_field}(a)). Different from Ref.~\cite{Schuster2010,Dykman2023-hx}, we propose to apply an external magnetic field normal to the liquid helium surface. The stray magnetic fields are created by the ferromagnet at the position of the Rydberg-ground state and the Rydberg-1st excited state, which are presented in Fig.~\ref{fig:imagecharge_wavefunctions}(c) and in Fig.~\ref{fig:imagecharge_wavefunctions}(d), respectively. The created magnetic field gradient introduces both the interaction between the Rydberg state and the spin state (Rydberg-spin interaction) and the spin-orbit interaction. By sending an AC voltage $\propto \cos(2 \pi f_\mathrm{0,spin} t)$ to one of the outer electrodes, one-qubit gates can be realized in an electric-dipole-spin-resonance (EDSR) manner~\cite{Tokura2006}. A sequence of MW pulses sent through the waveguide (Fig.~\ref{fig:electrode_dimension_slanting_magnetic_field}(b)) works as a controlled-phase gate for the spin state
with the assistance of the intrinsic electric dipole-dipole interaction and the artificial Rydberg-spin interaction. A quantum-non-demolition (QND) readout of the spin state can  be accomplished  by reading out the impedance change by sending an AC voltage $\propto \cos(2 \pi f_m t)$ to the LC circuit.

In this work, we consider the case where the helium depth is $140~$nm, which is thinner than the case considered in Ref.~\cite{Dykman2003}, where the helium depth is $500~$nm. In addition to this difference, here, we have segmented toroidal electrodes (outer electrodes). We need to use a thin helium film in order to obtain a large enough magnetic field gradient at the position of the electron. Additionally, the center electrode should be close enough to the electron to obtain a large enough image charge difference between the Rydberg ground state and the Rydberg first excited state. 

\section{Hamiltonian of the system\label{sec:hamiltonian}}

The electric potential felt by the electron $V(x,y,z)$  can be divided into three parts, that is $V(x,y,z)=V_\mathrm{ImageHe} +  V_\mathrm{ImageElectrodes} +  V_\mathrm{VoltageElectrodes},$ where  
$V_\mathrm{ImageHe}$ is the potential created by the electrostatic image of the electron in  liquid helium, $V_\mathrm{ImageElectrodes}$ is the potential created by the electrostatic image in the electrodes, and $V_\mathrm{VoltageElectrodes}$ is the potential created by the voltage applied to the electrodes. The helium depth $140$~nm is  high enough to neglect the image charge induced on the electrodes by the image charge in helium. Thus, we can treat the image charge in liquid helium and the image charge on the electrodes separately. The potential created by the electrostatic image in  liquid helium is written as $V_\mathrm{ImageHe}=\frac{\epsilon_\mathrm{He}-1}{\epsilon_\mathrm{He}+1} \frac{e}{16 \pi \epsilon_0 z},$ where $\epsilon_0$ is the electrical permittivity of vacuum and $\epsilon_\mathrm{He}$ is the relative permittivity of liquid helium-4. Differently from the case treated in Ref.~\cite{Dykman2003}, here, the helium depth $140$~nm is so small that $V_\mathrm{ImageElectrodes}$  and $V_\mathrm{VoltageElectrodes}$ cannot be written as a simple analytical expression or cannot be separated into the $z$ direction and the $x,y$ direction components. Instead, they were obtained numerically by solving Poisson's equation by the finite element method using COMSOL. For calculating $V_\mathrm{ImageElectrodes}$, we approximate an electron by a small spherical conductor whose surface charge is equal to the elementary charge and which is placed at an arbitrary position. In order to
eliminate the contribution to the electron energy from the electric potential due to the
electron itself, we subtract the calculated potential without electrodes from the calculated potential with all electrodes included.  To calculate $V_\mathrm{VoltageElectrodes}$, we separately calculate the potential due to each electrode by applying 1~V on a single electrode, while keeping all the other electrodes grounded and summing them together.

A static magnetic field $ B_0$ is applied normal to the helium surface, and therefore the electron is trapped both magnetically and electrically. Using the symmetric gauge, the vector potential reads $\bm{A}=\left(-\frac{B_0y}{2}, \frac{B_0x}{2}, 0\right)$. The energy-eigenstate equation with such a magnetic field can be written as
\begin{equation}
H_0\Phi (x,y,z)=E^{(0)} \Phi (x,y,z) \label{Hamiltonian_xyz}
\end{equation}
with\begin{align}
 H_0&=\frac{(p_x-\frac{eB}{2}y)^2}{2m_e}+\frac{(p_y+\frac{eB}{2}x)^2}{2m_e}-e V(x,y,z) \label{Hamiltonian1}\\
       &=\frac{p_x^2}{2m_e}+\frac{p_y^2}{2m_e}+\frac{1}{2}\omega_cL_z +\frac{1}{8} m_e \omega_c^2 r^2
       -e V(x,y,z),
\end{align}
where $\omega_c=eB_0/m_e$ is the cyclotron frequency,  $L_z=p_y x- p_x y$ and $r=\sqrt{x^2+y^2}$ (Fig.~\ref{fig:electrode_dimension_slanting_magnetic_field}(a)), 
$m_e$ is the electron mass, and $e>0$ is the elementary charge. In the calculations presented here, we assume the same DC voltage is applied to the two outer electrodes. The gap between the outer electrodes is sufficiently small compared to the liquid helium depth and ergo the DC electric potential is approximately azimuthally symmetric.  Therefore, the two outer electrodes can be approximated by a hollow cylinder of outer diameter  900~nm and of  inner diameter 300~nm (Fig.~\ref{fig:electrode_dimension_slanting_magnetic_field}(a)). 

Using the cylindrical coordinates with the ($r, \theta$) plane  parallel to the liquid surface, the electric potential felt by the electron and the orbital wavefunction can be written as $V(r,z)$ and  $\Phi(r,z,\theta)=\phi_{m, n_r, n_z}(r,z) \psi_m(\theta)$, respectively, since both the DC electric and magnetic potentials are approximately azimuthally symmetric. Using $\psi_m(\theta)=\frac{1}{\sqrt{2 \pi}} e^{i m\theta} $ with $i^2=-1$, $m=0,\pm 1,\pm 2, \pm3..$, the energy-eigenstate equation is rewritten as 
\begin{equation}
H_0\phi_{m, n_r, n_z}(r,z)=E^{(0)}_{m, n_r, n_z} \phi_{m, n_r, n_z}(r,z)\label{Shrodinger_rz},
\end{equation}
with
\begin{equation}
H_0= - \frac{\hbar^2}{2m_e} \left( \frac{1}{r}\frac{\partial}{\partial r}  +\frac{\partial^2}{\partial r^2}+\frac{\partial^2}{\partial z^2} -\frac{m^2}{r^2} \right) +\frac{1}{2}\hbar \omega_c m +\frac{1}{8} m_e  \omega_c^2 r^2   -e V(r,z) 
\label{Hamiltonian_0}.
\end{equation}

We solved Eq.~(\ref{Shrodinger_rz}) numerically with the electric potential $V(r,z)$ calculated as described earlier. Even though $V(r,z)$ cannot be separated into the in-plane ($r$) and the vertical ($z$) components, we can still label the eigenfunctions $\phi_{m, n_r, n_z}(r,z)$ by the quantum numbers $n_r$ and $n_z$ associated with the electron's in-plane orbital motion (orbital state) and vertical orbital motion (Rydberg state), respectively, with $n_r=0,1,2,3...$ and $n_z=1,2,3...$ 

\begin{figure}[H]
\centering
\includegraphics[width=\textwidth]{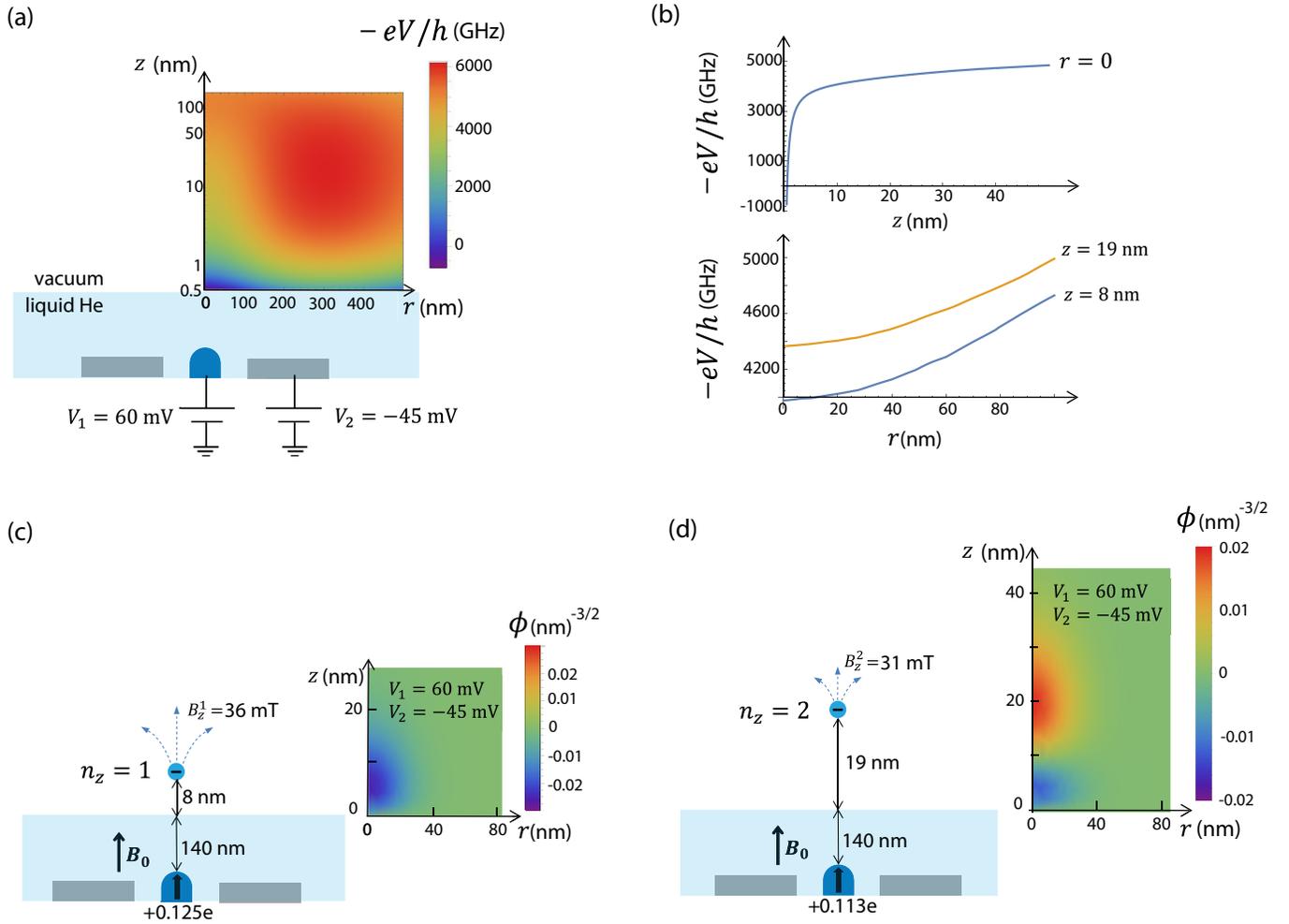}
\caption{\label{fig:imagecharge_wavefunctions} Focus on one electron. The cylindrical coordinate ($r,\theta, z$) is used here. The ($r, \theta$) plane is parallel to the liquid surface. The origins for $r$ and $z$ are set to the position of the electron and the liquid surface, respectively. (a) Electric potential energy of the electron, $-eV(r,z)/h$ with $V_1=60~$mV and $V_2=-45~$mV. A cut at $r=0$ along $z$ and cuts at $z=19$ nm and $z=8$ nm along $r$ are shown in (b). The z-axis is on a logarithmic scale for clarity. 
(c,d) When $V_1=60~$mV and $V_2=-45~$mV, the average position of the electron along the $z$ axis is 8~nm and 19~nm from the helium surface for the Rydberg-ground state ($n_z=1$) in (c), and for the Rydberg-1st-excited state ($n_z=2$) in (d), respectively. The insets show normalized eigenfunction $\phi(r,z)$ for $(m,n_r,n_z)=(0,0,1)$ in (c) and $=(0,0,2)$ in (d).  The image charge induced on the center electrode is $0.125e$ and $0.113e$ for  $n_z=1$ and  $n_z=2$, respectively. 
An external magnetic field $B_0$ is applied along the $z$ axis and the pillar is magnetized along the same axis. The field lines of the stray magnetic field are shown at the position of the electron for $n_z=1$ (for $n_z=2$) in (c) (in (d)). $B_z^1$ ($B_z^2$) is the $z$ component of the stray magnetic field felt by the electron in the Rydberg-ground (Rydberg-1st-excited) state. }
\end{figure}

To confine an electron to the central region, the central pillar is biased positively, while the outer electrodes are biased negatively. Concurrently, we should carefully adjust the voltages to avoid the pressing field experienced by the electron from becoming excessive, leading to a higher Rydberg transition energy.  In order to send a high enough power with relatively inexpensive equipment (WR4 waveguides and multiplier~\cite{VDI}), we strive to keep the Rydberg transition energy below 240~GHz. If the transition energy is too high, it is not straightforward to send a resonant MW of high enough power, and the MW multiplier becomes expensive~\cite{VDI}. Based on these requirements, we choose $V_1=60~$mV and $V_2=-45~$mV, such that the Rydberg transition energy is 240~GHz (Appendix~\ref{wavefunction_energy}) and the electron is well confined around the central region. 

Fig.~\ref{fig:imagecharge_wavefunctions}(a,b) consequently show the potential $V(r,z)$ under these voltage conditions, and the insets of Fig.~\ref{fig:imagecharge_wavefunctions}(c) and (d) depict the wavefunction of the electron in the Rydberg-ground state and in the Rydberg-1st-excited state, respectively. The electric potential along the $z$ axis, as seen in Fig.~\ref{fig:imagecharge_wavefunctions}(b), reveals a hydrogen-like potential, with an infinity barrier assumed at $z=0$. The electric potential along the $r$ axis is approximately harmonic, showing a slight dependence on the $z$ coordinate. The characteristic length of the dot size, calculated as $l_0=|\bra{\phi_{m=0,n_r=0,n_z=1}} r \ket{\phi_{m= \pm 1,n_r=0,n_z=1}}|$, is $17.8$~nm for the chosen voltage values.

The vertical and in-plane stray magnetic fields created by the ferromagnetic pillar near the position of the electron in the Rydberg-ground state can be approximately described as $b_z(r,z)=B_z^1+\frac{\partial b_z}{\partial z} (z-z_{11})+\frac{1}{2}\frac{\partial^2 b_z}{\partial r^2} r^2$ and $b_r(r,z)=\frac{\partial b_r}{\partial r} r$, respectively. $z_{n_zn_z'}$ stands for $\bra{\phi_{0,0,n_z}}z \ket{\phi_{0,0,n_z'}}$ and therefore $z_{11}$ is the average position of the electron in the Rydberg-ground state. $B_z^1$ and $\frac{\partial b_z}{\partial z}$, $\frac{\partial^2 b_z}{\partial r^2}$, and $\frac{\partial b_r}{\partial r}$ are the $z$ component of the stray field and the field gradients at $(r,z)=(0,z_{11})$, respectively. To account for the effect of the ferromagnetic pillar, we rewrite the Hamiltonian as $H=H_0+W + \frac{1}{2}g \mu_B ( B_0+ B_z^1)\sigma_z$, where $H_0$ is given by Eq.~(\ref{Hamiltonian_0}). $W=W_{zz}+W_{rr}+W_{rz}$ is the Hamiltonian  of the charge-spin interaction, where   

\begin{equation}
W_{zz}= 
    \frac{1}{2}g\mu_{\rm B}  \frac{\partial b_z}{\partial z}(z-z_
    {11})  \sigma_z
\label{W_zz}
\end{equation}  is the Hamiltonian of the Rydberg-spin interaction and $W_{rr}+W_{rz}$ is the Hamiltonian of the spin-orbit interaction with
\begin{equation}
W_{rr}= \frac{1}{2}g\mu_{\rm B}\frac{\partial b_r}{\partial r}r \sigma_r
\label{W_rr}
\end{equation} 
and
\begin{equation}
W_{rz}=\frac{1}{4} \frac{\partial ^2b_z}{\partial r^2}r^2    \sigma_z
\label{W_rz}.
\end{equation} Here, $\sigma_\square$ with $\square=x, y, z$  are Pauli matrices spanned by the spin-up and spin-down states,  $\sigma_r=\sigma_x \cos \theta + \sigma_y \sin \theta $, $g$ is the free electron Lande g factor, and  $\mu_\mathrm{B}$ is the Bohr magneton. 

\section{\label{sec:single-qubits} Single-qubit gates}

 \begin{figure}[!h]
\centering
\includegraphics[width=0.8\textwidth]{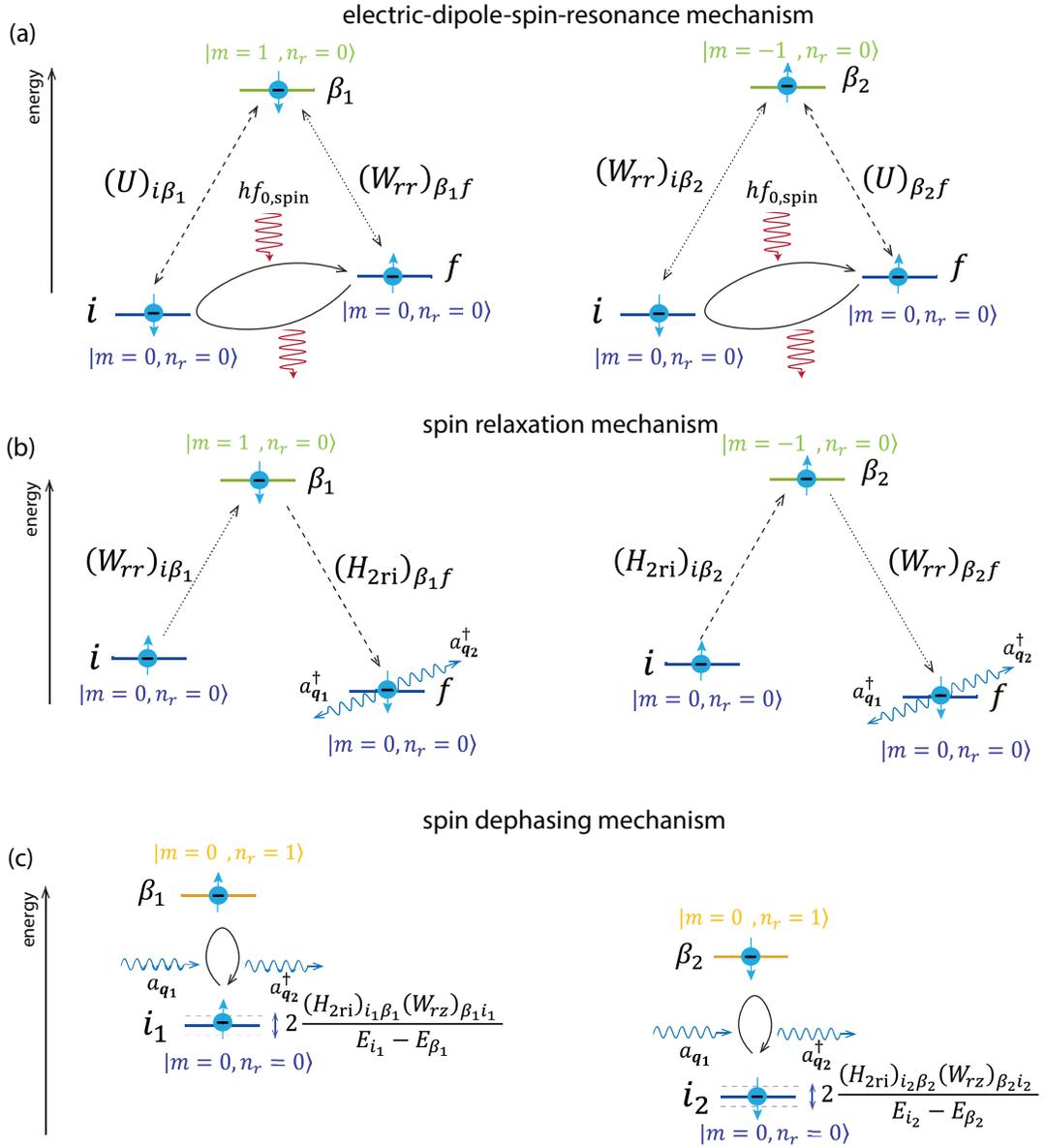} 
\caption{\label{fig:relaxation_dephasing_mechanism} 
See also the main text. (a) Two second-order processes of the EDSR.  The process shown on the left (right) panel corresponds to the first (second) term in $|...|$ of Eq.~(\ref{eq:SpinRabi_0}). The wavy red arrows represent absorbed and emitted photons. (b) Two second-order processes of the spin relaxation. The process shown on the left (right) panel corresponds to the first (second) term in $|...|^2$ of Eq.~(\ref{eq_Gamma_relax2}). The wavy blue arrows represent emitted ripplons. (a,b) $i$ and $f$ stand for the initial and final states, respectively, and $\beta_1$ and $\beta_2$ stand for the intermediate state of the second-order process.  The virtual transition with simultaneous flipping of the orbital state and the spin state (dotted arrows) can be caused due to the in-plane field gradient, with the corresponding Hamiltonian $W_{rr}$.  Dashed arrows represent the virtual transitions induced by the AC electric field with the corresponding Hamiltonian $U$ in (a) and  by the two-ripplon scattering with the corresponding Hamiltonian $H_{2\mathrm{ri}}$ in (b). As a result, the spin state is flipped for $m=0$ in the end. (c)  The fluctuation of the energy levels for the spin-up and spin-down states (the first and second terms of Eq.~(\ref{eq_delta_E_2}))
are depicted by the dark blue double arrows and the second-order processes that introduce the energy fluctuation are also shown on the left and right panels, respectively. $i_1$ and $i_2$ stand for the initial state of the spin-up state and the spin-down state, respectively. They are excited to the intermediate state by the second-order process: $\beta_1$ and $\beta_2$, respectively, and return back to the original states in the end. Different from (b), one ripplon is emitted and another ripplon of the same energy is absorbed (wavy blue arrows), and therefore the total energy change of the ripplon system is zero.}
\end{figure}
In the following discussion, we express the state as $\{m,n_r,n_z,\sigma \}$, where  $\sigma$ represents the spin-up state ($\sigma=\uparrow$ or +1) or the spin-down state  ($\sigma=\downarrow$ or -1) and we omit some indexes for clarity when the electron is in the ground state for those quantum numbers.

 Single-qubit gates for the spin state can be realized by applying an AC voltage to one of the two outer electrodes and thus creating an AC electric field along the $x$ axis at the position of the electron with the corresponding Hamiltonian for the electron interaction with the electric field $eE^\mathrm{RF} x\cos(2 \pi f_\mathrm{0,spin}t)$, where  $hf_{\mathrm{0,spin}} \approx g \mu_B B_0$ is the energy splitting between the state  $i=\{m=0, \downarrow\}$ and the state $f=\{m=0,\uparrow \}$. This electric field interacts with the orbital degree of freedom and due to the spin-orbit interaction created by the magnetic field gradient,  we can control the spin state coherently in an EDSR manner \cite{Tokura2006}.  The part of the Hamiltonian which is responsible for the coherent control of the spin state is given by 
 \begin{equation}
    H(t)=eE^\mathrm{RF} x\cos(2 \pi f_\mathrm{0,spin}t)+W. \label{eq:Ht_onequbit}
 \end{equation} 
 The spin-flip transition happens due to second-order processes, virtual transitions of an electron to excited orbital states (Fig.~\ref{fig:relaxation_dephasing_mechanism}(a)). The off-diagonal effective Hamiltonian term that induces the resonance transition between  $i=\{m=0, \downarrow\}$ and  $f=\{m=0,\uparrow \}$ is given by~\cite{Winkler2003,Romhanyi2015}  (Appendix~\ref{sec:coherent_control}) 
\begin{equation}
\frac{1}{2}   \sum_{\beta \neq i,f} (H(t))_{ i \beta} (H(t))_{ \beta f}
\left( \frac{1}{E_i-E_{\beta}} + \frac{1}{E_f-E_{\beta}} \right),
\end{equation}
where $(H(t))_{\triangle, \bigcirc}$  stands for $\bra{\triangle} H(t)\ket{\bigcirc}$ and $\sum_{\beta \neq i,f}$ refers to the sum over all the  eigenstates.   
$E_\square$ stands for the electron energy of the state $\square=\{m,n_r,n_z,\sigma \}$, where  $E_{m,n_r,n_z,\sigma}=E^{(0)}_{m,n_r,n_z}+\frac{1}{2} g \mu_\mathrm{B} B_0 \sigma$. Considering only the two lowest excited states that give non-zero terms: $\beta_1=\{m=+1,\downarrow \}$ and $\beta_2=\{m=-1,\uparrow \}$ as the eigenstates, the Rabi frequency rate is approximated to be 

\begin{equation}
f_\mathrm{1,spin}=\frac{1}{2 \pi \hbar} \frac{1}{2}  \left|   
(U)_{i \beta_1} (W_{rr})_{\beta_1 f} \left( \frac{1}{E_i-E_{\beta_1}} + \frac{1}{E_f-E_{\beta_1}} \right)
+
(W_{rr})_{ i \beta_2} (U)_{\beta_2 f } \left( \frac{1}{E_i-E_{\beta_2}} + \frac{1}{E_f-E_{\beta_2}} \right)
\right|,
\label{eq:SpinRabi_0}
\end{equation}
where $(W)_{\triangle, \bigcirc}$  stands for $\bra{\triangle} W \ket{\bigcirc}$ and $(U)_{\triangle, \bigcirc}$  stands for $ \bra{\triangle} U \ket{\bigcirc}$ with $U=e E^\mathrm{RF}x/2$. Note that the contributions from  $W_{zz}$ and $W_{rz}$ are zero and the virtual spin-flip transitions induced by the spin-orbit interaction $W_{rr}$ require a change in the orbital magnetic number $m$ by $\pm 1$ due to the conservation of angular momentum. Furthermore, the contributions from the higher states ($n_r \geq 1$ and $n_z \geq 2$) are also negligibly small.  From this, we obtain the Rabi frequency of the spin state,  

\begin{equation}
    f_\mathrm{1,spin}=\frac{1}{8h}  \partial_r f_{b_r}e E^\mathrm{RF} l_0^2  \frac{1}{2} \left|\sum_{\beta=\beta_1,\beta_2}   \left(  \frac{1}{f_i-f_{\beta}}+
\frac{1}{f_f-f_{\beta}} \right) \right|,
\label{eq:SpinRabi}
\end{equation}
where $\partial_r f_{b_r}=g \mu_B \frac{\partial b_r}{\partial r}/h$ and $f_\square=E_\square/h$  for $\square=i,\beta$ . Our numerical simulation shows that the in-plane magnetic field gradient is $\frac{\partial b_r}{\partial r}=0.23$~mT/nm when the cobalt pillar is fully magnetized along the $z$ axis after reaching its saturation magnetization (1.8~T). With this voltage configuration,  $f_\mathrm{0,spin} \approx \omega_c/2 \pi=16$~GHz, and the strength of the electric field at the position of the electron, $E^\mathrm{RF}=500$~V/cm, the Rabi frequency of the spin state is calculated to be $f_\mathrm{1,spin}\approx 100~$MHz (Appendix~\ref{sec:single-qubits}).  $E^\mathrm{RF}=500$~V/cm can be realized by applying an AC voltage to one of the outer electrodes (Fig.~\ref{fig:electrode_dimension_slanting_magnetic_field}(a)) with an amplitude  $\alpha E^{\mathrm{RF}}=$12.5~mV, which is small enough not to cause a significant heating problem experimentally. From a COMSOL simulation, the effective distance for the voltage applied to the outer electrode to determine the electric field at the position of the electron is calculated to be  $\alpha = 250 $~nm. The generic single-qubit rotation gates can be realized by adjusting the phase of the AC voltage. Since single-qubit gates are performed by sending an AC voltage to each electrode allocated to each qubit, qubits can be individually addressed.

When single-qubit gates are performed and calculation is idle, the qubit states are always in the Rydberg-ground state therefore we do not directly suffer from the fast relaxation of the Rydberg state. However, the introduced spin-orbit interaction and Rydberg-spin interaction open a path for the spin state to relax or dephase via the relaxation or the dephasing of the orbital state and the Rydberg state.  The spin relaxation and dephasing rates are calculated in Sec.~\ref{sec:spin_relaxation} and Sec.~\ref{sec:spin-dephasing}. The single-qubit gate fidelity is determined by  $f_\mathrm{1,spin}\approx 100~$MHz  and the spin relaxation time $T_\mathrm{1,spin}
  \approx 50~$ms (Sec.~\ref{sec:spin_relaxation}) and it is calculated to be $F_1>99.9999\%$ (Appendix~\ref{sec:single_qubit_gate_fidelity}).

\section{Spin relaxation \label{sec:spin_relaxation}}

While the spin-orbit interaction allows us to realize single-qubit gates, it also opens the path for the spin-up state to relax to the spin-down state. Such relaxation is induced by the virtual transitions between electron orbital states that happen due to the interaction between the electron and the liquid helium surface capillary waves called ripplons (Fig.~\ref{fig:relaxation_dephasing_mechanism}(b)). The spin relaxation is the energy non-conserving process and is dominated by the emission of two short-wavelength ripplons \cite{Monarkha2007-el} (Appendix~\ref{sec:longitudinal_relaxation_appendix}). When the combined energy of the emitted two ripplons matches the Zeeman energy $h f_\mathrm{0,spin}$, the spin relaxation can occur in the second-order process.  The Hamiltonian of the electron-ripplon interaction, which corresponds to the two-ripplon processes, is quadratic in surface displacement and is given by 

\begin{equation}
   H_{\mathrm{2ri}}=\sum_{\bm{q_1}} \sum_{\bm{q_2}} \xi_{\bm{q_1}} 
\xi_{\bm{q_2}} 
U_{\bm{q_1}\bm{q_2}} e^{i \bm{q_1} \bm{r}} e^{i \bm{q_2} \bm{r}}, \label{eq_H2ri}  
\end{equation}
where 
$\xi_{\bm{q}}=Q_{\bm{q}}( a_{\bm{q}}+ a_{-\bm{q}}^\dagger) $
is the Fourier transform of the operator of the surface displacement of liquid helium, $a_{\bm{q}}$ and $a_{\bm{q}}^\dagger$ are the  creation and annihilation operators for ripplons, respectively, $Q_{\bm{q}}=\sqrt{\frac{\hbar q}{2S \rho \omega_q}}$, $S$ is the surface area of the liquid and $\rho$ is the density of liquid helium. $U_{\bm{q_1}\bm{q_2}}$ is the electron-two-ripplon coupling  (Appendix~\ref{sec:longitudinal_relaxation_appendix}). Here, we take $H_\mathrm{2ri}+W$ as the perturbation term. According to Fermi's golden rule in the second-order perturbation theory, the relaxation rate is given by 
\begin{equation}
T_\mathrm{1,spin}^{-1}=\frac{2\pi}{\hbar} \left|  \sum_{\beta \neq i,f}
\frac{ (W)_{ i\beta} (H_\mathrm{2ri})_{\beta f }  }{E_i-E_{\beta}}+
\frac{ (H_{\mathrm{2ri}})_{i \beta } (W)_{\beta f }  }{E_i-E_{\beta}+\epsilon_{n_\mathrm{ri}}-\epsilon_{n_\mathrm{ri}'}}
\right|^2\delta(E_f-E_i+\epsilon_{n_\mathrm{ri}'}-\epsilon_{n_\mathrm{ri}}),
\label{Gamma_relax}
\end{equation}
where $i=\{m=0, \uparrow\}$, $f=\{m=0,\downarrow\}$, and $\sum_\beta$ refers to the sum over all the eigenstates. $(H_\mathrm{2ri})_{\triangle, \bigcirc}$  stands for $ \bra{\triangle, {n_\mathrm{ri}}} H_\mathrm{2ri} \ket{\bigcirc, {n'_\mathrm{ri}}}$, where $n_\mathrm{ri}$ and $n_\mathrm{ri}'$ collectively stand for the states of all the ripplons before and after the two-ripplon emission, respectively.  $\epsilon_{n_\mathrm{ri}}$ and $\epsilon_{n'_\mathrm{ri}}$ are the energies of the collective states of  the  ripplons  before  and  after  the  two-ripplon  emission, respectively, and therefore $\epsilon_{n_\mathrm{ri}'}-\epsilon_{n_\mathrm{ri}}=\hbar \omega_{q_1}+\hbar \omega_{q_2}$, where $\hbar \omega_{q_1}$and $\hbar \omega_{q_2}$ are the energies of the emitted ripplons \cite{Dykman2003}.  In the same way as for $f_\mathrm{1,spin}$ in Eq.~(\ref{eq:SpinRabi_0}),  considering only the two lowest excited states: $\beta_1=\{m=+1,\downarrow \}$ and $\beta_2= \{m=-1,\uparrow \}$,  the relaxation rate is approximated to be 

\begin{equation}
T_\mathrm{1,spin}^{-1}\approx \frac{2\pi}{\hbar} \left| 
\frac{(W_{rr})_{i \beta_1 } (H_\mathrm{2ri})_{\beta_1 f }   }{E_i-E_{\beta_1}}+ 
\frac{ (H_{\mathrm{2ri}})_{i \beta_2} (W_{rr})_{ \beta_2 f} }{E_i-E_{\beta_2}+\epsilon_{n_\mathrm{ri}}-\epsilon_{n_\mathrm{ri}'}}
\right|^2\delta(E_f-E_i+\epsilon_{n_\mathrm{ri}'}-\epsilon_{n_\mathrm{ri}})
\label{eq_Gamma_relax2}
\end{equation}
(Fig.~\ref{fig:relaxation_dephasing_mechanism}(b)).  With $V_1=60~$mV,  $V_2=-45~$mV,  $f_\mathrm{0,spin} \approx \omega_c/ 2 \pi  =16$~GHz, the spin  relaxation time $T_\mathrm{1,spin}$ is calculated to be $\approx 50~$ms (Appendix~\ref{sec:longitudinal_relaxation_appendix}). 

In Ref.~\cite{Dykman2023-hx}, the energy difference between the spin-up orbital-ground state and the spin-down orbital-1st-excited state is set to be small to make the one-qubit gate fast enough. This small energy detuning opens a path for quasi-elastic spin-orbit-flip process induced by an absorption or emission of a single ripplon, thus significantly increasing the spin relaxation rate. For the energy detuning 5~MHz, they estimated $T_\mathrm{1,spin}^{-1}=$ $0.6 \times 10^3$~s$^{-1}$ and $6 \times 10^5$~s$^{-1}$ for the pressing field $E_\perp$=0 and 300 V/cm, respectively~\cite{Dykman2023-hx},  which is much faster than our estimation:  $T_\mathrm{1,spin}^{-1}=$$20$~Hz for  $E_\perp=200$~V/cm. In this proposal, thanks to a higher magnetic field gradient than in Ref.~\cite{Dykman2023-hx}, we can have fast enough single-qubit gates and two-qubit gates even if the spin-up orbital-ground state and the spin-down orbital-1st-excited state are highly detuned.

\section{\label{sec:spin-dephasing}Spin dephasing}

The dephasing of the spin state happens due to the fluctuation of the Zeeman splitting over time. Such a fluctuation is induced by spin-conserving virtual transitions of an electron to higher orbital states. Similar to the dephasing of the Rydberg states introduced in \cite{Dykman2003}, the dephasing time of the spin state $T_\mathrm{2,spin}$ is related to the auto-correlation function of 
 the time-dependent fluctuation of the Zeeman energy $\delta E(t)  $: $\left< \delta E(t')\delta E(t'')  \right>=\hbar^2  \delta(t'-t'')/T_\mathrm{2,spin}$, where $\delta(t)$ is the Dirac delta function.  Taking $H_\mathrm{2ri}(t)+W$ as the perturbation term (see Appendix~\ref{sec:spindephasing_appendix} for $H_\mathrm{2ri}(t)$), the time-dependent energy is calculated as the second-order energy shift caused by the perturbation term:
  \begin{equation}
  \delta E(t)= \sum_{\beta \neq i_1}\frac{ (H_\mathrm{2ri}(t) +W)_{i_1 \beta} (H_\mathrm{2ri}(t) +W  )_{\beta i_1}}{E_{i_1}-E_\beta}  -  \sum_{\beta' \neq i_2 }\frac{ (H_\mathrm{2ri}(t) +W)_{i_2 \beta'} (H_\mathrm{2ri}(t) +W)_{\beta' i_2} }{E_{i_2}-E_{\beta'}} ,
 \end{equation}
  where $i_1=\{n_r=0, \uparrow\}$, $i_2=\{n_r=0,\downarrow \}$. The virtual transitions to $m\neq 0$ states are forbidden because of the conservation of the electron angular momentum. Considering only the two lowest excited states: $\beta_1=\{n_r=1,\uparrow \}$ and $\beta_2=\{n_r=1,\downarrow\}$, as the contribution from the higher excited states ($n_r \geq 2$ and $n_z \geq 2$) is negligibly small, the time-dependent Zeeman energy fluctuation is approximated to be 
  \begin{equation}
 \delta E(t)\approx 2 \frac{ (H_\mathrm{2ri}(t))_{i_1 \beta_1 }  (W_{rz})_ {\beta_1 i_1 } }{E_{i_1}-E_{\beta_1}} -  2\frac{ (H_\mathrm{2ri}(t))_{ i_2 \beta_2} 
 (W_{rz})_{ \beta_2 i_2} }{E_{i_2}-E_{\beta_2}} \label{eq_delta_E_2}
 \end{equation}
(Fig.~\ref{fig:relaxation_dephasing_mechanism}(c)).  The virtual orbital transitions between $n_r=0$ to $n_r=1$ states are induced by the two-ripplon scattering of the thermally excited long-wavelength ripplons (Appendix~\ref{sec:spindephasing_appendix}) mediated by a magnetic field gradient. As the contribution from the $W_{zz}$ is negligibly small and that from $W_{rr}$ is zero, the second-order in-plane field gradient $\frac{\partial^2 b_z}{\partial r^2}$ in $W_{rz}$ in Eq.~(\ref{W_rz}) is responsible for the spin dephasing. With $V_1=60~$mV,  $V_2=-45~$mV,  $f_\mathrm{0,spin}\approx  \omega_c/ 2 \pi  =16$~GHz, and the liquid helium temperature $T=100~$mK, the spin dephasing time $T_\mathrm{2,spin} $ is calculated to be $> 100~$s (Appendix~\ref{sec:spindephasing_appendix}). As the process is dominated by the thermally excited ripplons, the dephasing time becomes even longer as the liquid helium temperature is lowered (Appendix~\ref{sec:spindephasing_appendix}). This dephasing time is much longer than the relaxation time calculated in  Sec.~\ref{sec:spin_relaxation} and thus the coherence time is determined by the relaxation time.

\section{\label{sec:two-qubit}two-qubit gate}
A two-qubit gate for the spin state can be realized by exciting the Rydberg state spin-selectively and depending on the Rydberg state of an adjacent electron. In Sec.~\ref{sec:Perturbation}, we show that the Rydberg transition energy of an electron depends on the spin state of the electron itself and the Rydberg state of an adjacent electron. In Sec.~\ref{sec:cphase}, a concrete MW pulse sequence to realize a controlled-phase gate is presented.

\subsection{Electric dipole-dipole interaction}\label{sec:Perturbation}

\begin{figure}
\centering
\includegraphics[width=1\textwidth]{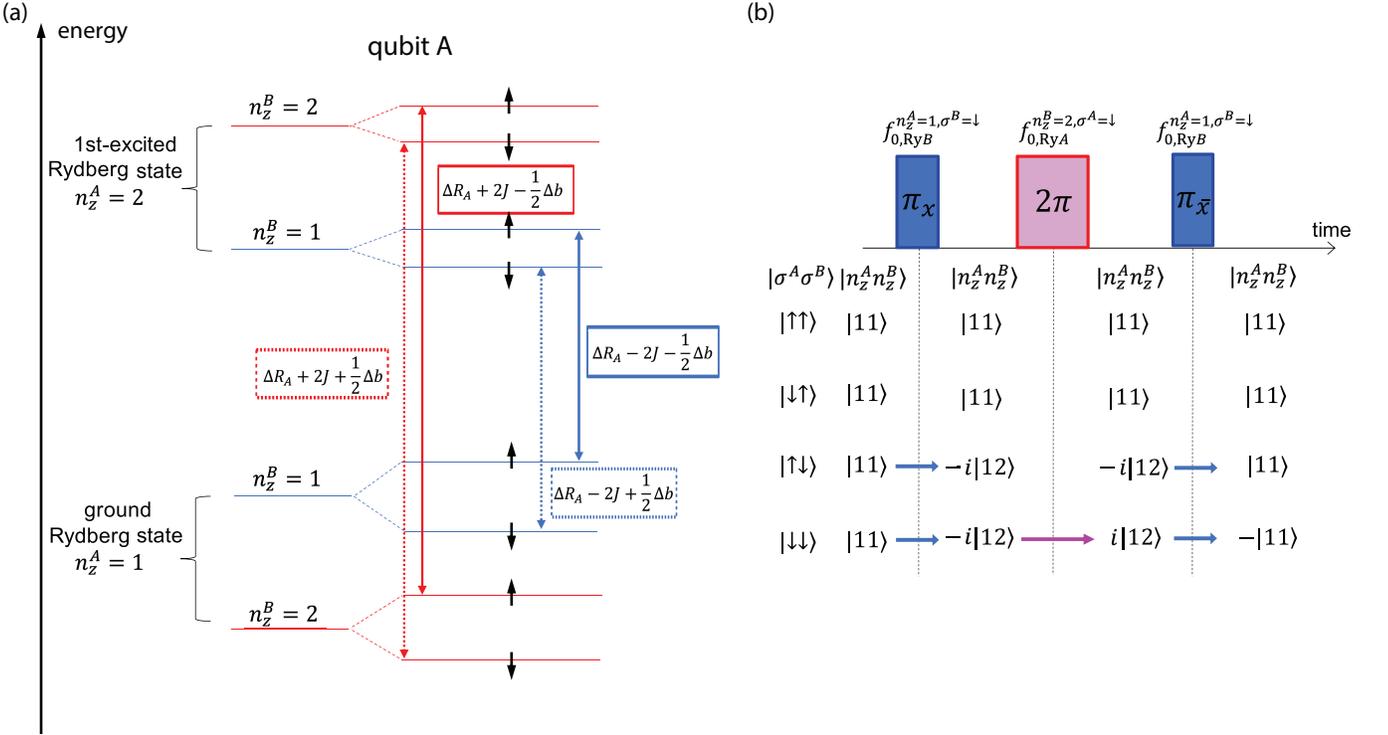}
\caption{\label{fig:energy_diagram_and_pulses} (a) Energy levels of electron A for the Rydberg state of the neighboring electron $n_z^B=1$ (shown in blue) and $n_z^B=2$ (shown in red). The transition energy of the Rydberg state of electron A (shown with dotted arrows for $\sigma^A=-1$ and with solid arrows for $\sigma^A=1$) depends on both the spin state of electron A and the Rydberg state of electron B. (b) MW pulse sequence to realize a controlled-phase gate. Here, we consider four possible combinations of the spin states of qubits A and B. The first and third pulses are effective only for the last two cases and the 2nd pulse is effective only for the last case. See the texts for the details.}
\end{figure}

When electrons are trapped close enough to each other, we should account for a Hamiltonian term corresponding to the Coulomb interaction, which depends on the distance between the electrons and thus depends on whether they are in the same Rydberg state. In the two-lowest level approximation, taking the Rydberg-ground state $\ket{n_z=1}$ and the Rydberg-1st-excited state $\ket{n_z=2}$ as basis states, the Hamiltonian of the electric dipole-dipole interaction can be expressed as 
\begin{align}
W'=\frac{e^2}{4 \pi \epsilon_0 d^3} \left( (z_{11}^A-z_{22}^A)(z_{11}^B-z_{22}^B)\frac{s_z^A s_z^B}{4} + 2z_{12}^A z_{12}^B \frac{s_x^A s_x^B + s_y^A s_y^B}{4} \right),
\label{Hamiltonian1}
\end{align} 
where $s_z^i=\ket{n_z^i=2}\bra{n_z^i=2}-\ket{n_z^i=1}\bra{n_z^i=1}$, $s_x^i=\ket{x+}\bra{x+}-\ket{x-}\bra{x-}$, $s_y^i=\ket{y+}\bra{y+}-\ket{y-}\bra{y-}$, $\ket{x\pm}=\frac{1}{\sqrt{2}} (\ket{n_z^i=2}\pm \ket{n_z^i=1})$, and $\ket{y \pm}=\frac{1}{\sqrt{2}} (\ket{n_z^i=2}\pm i\ket{n_z^i=1})$ for qubit A ($i=A$) or qubit B ($i=B$)~\cite{Dykman2003}. $\epsilon_0$ is the vacuum permittivity and $d$ is the in-plane distance between pillars. Treating $W'$ as the perturbation term, the first-order energy shift is given by
\begin{equation}
\bra{n_z^A} \bra{n_z^B }  W' \ket{n_z^A} \ket{n_z^B } 
\approx J (-1)^{n_z^A} (-1)^{n_z^B}
\label{energyshift}
\end{equation}
for $n_z^i=1$ or $2$. The interaction energy is $4J= \kappa \frac{e^2}{4 \pi \epsilon_0} \frac{(z_{11}^A-z_{22}^A)(z_{11}^B-z_{22}^B)}{d^3}$, where $\kappa$ is the factor acquired due to the screening effect (Sec.~\ref
{sec:screening}). Due to the Ryberg-spin interaction, the Zeeman energy for the Rydberg-ground state is larger than that for the Rydberg-1st-excited state by $\Delta b= g\mu_\mathrm{B} |\Delta B_z|$, with $\Delta B_z = B_z^2-B_z^1$ (Fig.~\ref{fig:imagecharge_wavefunctions}(c,d)). In the same way, the Rydberg transition energy also differs by $\Delta b$ depending on the spin state (Fig.~\ref{fig:energy_diagram_and_pulses}(a)). Note that the external magnetic field is much larger than the stray magnetic field and thus only the magnetic field along the quantization axis (the $z$ axis) is considered. In total, the first-order energy shift of the two-electron system caused by both the Coulomb interaction and  the magnetic field gradient is
\begin{align}
\sum_{i=A,B} \frac{1}{2} \Delta b (1-n_z^i) \sigma^i  +J (-1)^{n_z^A} (-1)^{n_z^B} .
\label{energyshift_total}
\end{align}
As a result, the energy for the transition between the Rydberg-ground state and the Rydberg-1st-excited state of qubit A (B) depends on the Rydberg state of qubit B (A) and the spin state of qubit A (B) (Fig.~
\ref{fig:energy_diagram_and_pulses}(a)). Here, the energy shift due to the magnetic dipole-dipole interaction, which is much smaller than the terms in Eq.~(\ref{energyshift_total}), is neglected.  The energy shift due to the band-like structure when qubits are placed in a triangular array is also negligibly small when the distance between neighboring electrons is much larger than the distance between the electron and the electrode underneath it~\cite{Dykman2003}.
 With $d=0.88~\mu$m and $z_{11}^A-z_{22}^A=z_{11}^B-z_{22}^B=11.5~$nm, we obtain $4J/h= 137~$MHz and  $\Delta b= g\mu_\mathrm{B} |\Delta B_z|\approx 4J$ with $\Delta B_z \approx -5~$mT.

\subsection{\label{sec:cphase} Pulse sequence for a CPhase gate}

Taking into account the first-order energy shift due to $W+W'$, the energy difference between the Rydberg-ground state and the Rydberg-1st-excited state of qubit $i$ can be explicitly rewritten as 
\begin{equation}
\Delta E_{n_z^i=1 \rightarrow 2} =\Delta R_i + 2J(-1)^{n_z^j} - \frac{1}{2} \Delta b \sigma^i,\label{energy_differene}
\end{equation}where $\Delta R_i$ is the energy difference between the Rydberg-ground state and the Rydberg-1st-excited state of qubit $i$  given by $H_0$ (Eq.~(\ref{Hamiltonian_0})) for a given combination of  $V_1$ and $V_2$. $n_z^j$ is the Rydberg state of an adjacent qubit $j$. The transition frequency for the Rydberg state of qubit $i$, $f_\mathrm{0,Ry}=\Delta E_{n^i_z=1 \rightarrow 2}/h$, depends on the three factors: qubit $i$'s spin state $\sigma^i$, the Rydberg state of the adjacent qubit $n^j_z$,  and the voltage applied to the electrodes. A MW signal is transmitted via a rectangular waveguide \cite{Lea2000} to excite the Rydberg state (Fig.~\ref{fig:electrode_dimension_slanting_magnetic_field}(b)). Although MW is applied to all the qubits globally through the waveguide, the individual addressing of qubits is realized by tuning the Rydberg transition energy by tuning $V_1$ and $V_2$. 

Note that the electrons always stay in the Rydberg-ground state except when we perform a two-qubit gate; thus  the electric dipole-dipole interaction $W'$ does not affect the qubit states as long as both qubits A and B are in the same Rydberg state: $ [W',s^A_j s^B_j]=0$ ($j=x,y,z)$. On the other hand, the magnetic dipole-dipole interaction alters the qubit states, however, the evolution due to the magnetic dipole-dipole interaction is orders of magnitude slower than any quantum operation shown here~\cite{Lyon2006},  and  thus can be neglected.

In the following, we show one way to realize a two-qubit gate, a Cirac-Zoller-type controlled-phase gate~\cite{Cirac1995-am}. Both qubits A and B are in the Rydberg-ground state before starting the controlled-phase gate. Therefore, the initial qubit state is written as
\begin{equation}
a\ket{11 \uparrow \uparrow}+b\ket{11 \downarrow \uparrow}+c\ket{1 1\uparrow \downarrow }+d\ket{1 1  \downarrow  \downarrow }, \label{initial_state_Cphase}
\end{equation}
where  $\ket{n_z^A n_z^B \sigma^A\sigma^B}$ is an abbreviation of  $\ket{n_z^A }_A \ket{n_z^B }_B \ket{ \sigma^A}_A \ket{\sigma^B}_B $ and  $|a|^2+|b|^2+|c|^2+|d|^2=1$. Fig.~\ref{fig:energy_diagram_and_pulses}(b) presents a MW pulse sequence to realize a controlled-phase gate for qubits A and B. First, we apply a $\pi$ pulse around the $x$ axis with MW frequency $f_\mathrm{MW}=f_{0,\mathrm{Ry}B}^{n_z^A=1,\sigma^B=\downarrow}= (\Delta R_B  - 2J +\frac{1}{2}\Delta b)/h$. The Rydberg state of qubit B is excited when the Rydberg state of qubit A is the ground state ($n_z^A=1$) and the spin state of qubit B is spin-down ($\sigma^B=-1$). Consequently, the qubit state becomes
\begin{equation}
 a\ket{ 1 1 \uparrow\uparrow}+ b\ket{1 1 \downarrow \uparrow}-i c\ket{12 \uparrow \downarrow }-i d\ket{12  \downarrow   \downarrow}. \label{state_Cphase_2}
\end{equation}
Second, we apply a $2\pi$ pulse around the $x$ axis with MW frequency $f_\mathrm{MW}=f_{\mathrm{0, RyA}}^{n_z^B=2,\sigma^A=\downarrow}= (\Delta R_A + 2J + \frac{1}{2}\Delta b)/h$. Thus, the qubit state becomes
\begin{equation}
a\ket{ 11 \uparrow \uparrow} +b\ket{11 \downarrow \uparrow}-i c\ket{12\uparrow \downarrow }+i d\ket{12 \downarrow  \downarrow }. \label{state_Cphase_3}
\end{equation}
Third, we apply a $\pi$ pulse around $-x$ axis with the same MW frequency as the first pulse. Consequently, the qubit state becomes
\begin{equation}
a\ket{11\uparrow \uparrow}+b\ket{11 \downarrow \uparrow}+c\ket{1 1\uparrow \downarrow }-d\ket{1 1  \downarrow  \downarrow }. \label{final_state_Cphase}
\end{equation}
Compared to the initial qubit state (Eq.~(\ref{initial_state_Cphase})), the sign of the phase is changed only when the initial state of both qubits A and B is spin-down. In total, a controlled-phase gate for the spin states has been realized by the three MW pulses.

The Rydberg transition rate can be tuned by the power of the MW sent through a waveguide (Fig.~\ref{fig:electrode_dimension_slanting_magnetic_field}(b)).  The Hamiltonian for the electron interaction with the vertical electric field of the MW is $eE^\mathrm{MW}\cos(2\pi f_\mathrm{MW} t)z$ and the transition rate, which determines the two-qubit gate speed, is expressed as
\begin{equation}
f_\mathrm{1,Ry}=\bra{ n_z^i=1,\sigma^i  } e E^\mathrm{MW} z \ket{n_z^i=2,\sigma^i } /2h= e E^\mathrm{MW} z_{12} /2h \label{Rydbergtransitionrate},
\end{equation}
and can reach $50~$MHz~\cite{Kawakami2021} with $E^\mathrm{MW}=1$~V/cm (which corresponds to about 2 $\mu$W of EM wave power through a 1~mm$^2$  rectangular waveguide) and $z_{12} =4$~nm. We should carefully tune the Rydberg transition energy of the qubits so that only the Rydberg transition of the target qubit in intended states is on resonance. 

\begin{figure*}[!h]
\centering
\includegraphics[width=0.5\textwidth]{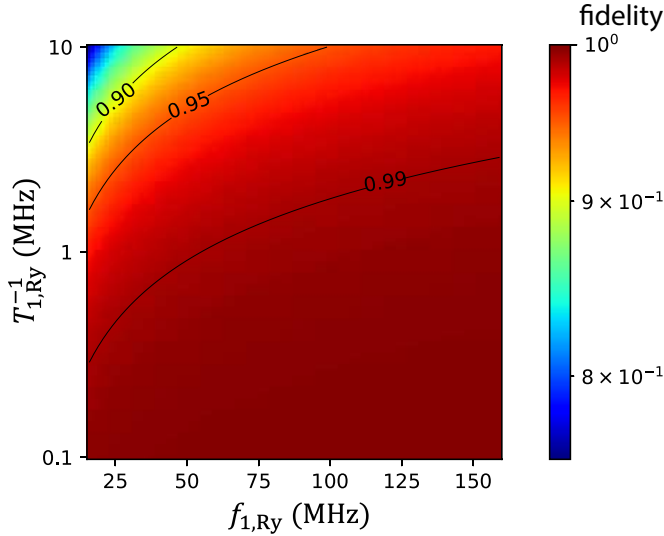}
\caption{\label{fig:two-qubit-gate_fidelity}  Fidelity of the controlled-($1.1885 \pi$) gate realized by applying the three MW pulses (Fig.~\ref{fig:energy_diagram_and_pulses} (b), Fig.~\ref{fig:energy_diagram_and_pulses_supplem}) with $\Delta b/h=4J/h=\sqrt{15} f_\mathrm{1,Ry}$.}
\end{figure*}

In fact, the target qubit in unintended states also acquires some phase during the Rydberg excitation through the Rydberg-spin interaction and the electric dipole-dipole interaction. We show how this effect can be compensated in detail in Appendix~\ref{sec:two_qubit_gate_appendix}. In short, a controlled-($1.1885\pi$) gate can be realized by the pulse sequence presented in Fig.~\ref{fig:energy_diagram_and_pulses} under the condition of $\Delta b/h=4J/h=\sqrt{15} f_\mathrm{1,Ry}$. For $\Delta b/h=4J/h=\sqrt{15}f_\mathrm{1,Ry}=137~$MHz (i.e., $f_\mathrm{1,Ry}=35.5$~MHz),  we obtained the gate fidelity for the controlled-($1.1885\pi$) gate $F_2=98.5\%$ (Appendix~\ref{sec:two-qubit_gate_fidelity_appendix}). As seen in Appendix~\ref{sec:controlled-pi}, the controlled-($\pi$) gate can be realized by applying the controlled-($1.1885\pi$) gate three times and single-qubit gates several times. The single-qubit gate fidelity is much higher than the two-qubit gate fidelity and therefore the controlled-($\pi$) gate fidelity is dominated by the controlled-($1.1885\pi$) gate fidelity and thus is calculated by $F_2^3$.  Fig.~\ref{fig:two-qubit-gate_fidelity} shows the fidelity of the  controlled-($1.1885\pi$) gate for different Rabi frequencies $f_\mathrm{1,Ry}$ and relaxation rates $T_\mathrm{1,Ry}^{-1}$ of the Rydberg state with $\Delta b/h=4J/h=\sqrt{15} f_\mathrm{1,Ry}$. We cannot simply increase $f_\mathrm{1,Ry}$ because of the constraint $\Delta b/h=4J/h=\sqrt{15} f_\mathrm{1,Ry}$ but should increase the field gradient and the Coulomb interaction at the same time. The field gradient can be increased by choosing a different ferromagnetic material that has a higher saturation magnetic field than Co, such as CoFe ($\approx 5\%$ higher)~\cite{Lachance-Quirion2015-yz} or Gd ($\approx 40\%$ higher)~\cite{Bertelli2015-fa}. Alternatively, we can make  the helium depth thinner but the Rydberg transition energy becomes higher. Sending a higher-frequency MW is more costly and difficult. The Coulomb interaction can be made larger by making the distance between electrons smaller. There are some limits on the electron density set by  the hydrodynamic instability but the distance between electrons can be made as small as $0.1~\mu$m~\cite{Marty1986-om}. $T_\mathrm{1,Ry}^{-1}$ was theoretically estimated and experimentally measured to be $\approx 1~$MHz for a liquid helium temperature lower than $100~$mK~\cite{Kawakami2021,Monarkha2010DecayRipplons,Monarkha2006-gi}. According to Ref.~\cite{Monarkha2010DecayRipplons,Monarkha2006-gi}, $T_\mathrm{1,Ry}^{-1}$ is limited by the two-ripplon emission of short-wavelength capillary waves and the entire discussion in this manuscript follows this theory, unless otherwise stated. However, other theoretical treatments of the two-ripplon scattering predict $T_\mathrm{1,Ry}^{-1} \le 0.1~$MHz~\cite{Dykman2003}.  With $T_\mathrm{1,Ry}^{-1} =0.1~$MHz (0.01~MHz) and $f_\mathrm{1,Ry}=35.5$~MHz, the controlled-($\pi$) gate (CZ gate) fidelity of $\approx 99.3\%$ ($\approx 99.7\%$) can be achieved. 

The proposed pulse sequence is one of the ways to realize a two-qubit gate. For example, making use of the Rydberg-blockade or Rydberg-antiblockade effect may allow us to realize a two-qubit gate with fewer steps or with a shorter time~\cite{Saffman2010-bb,Su2016-hg,Su2017-yg,Shi2017-gy}. Another improvement may be achieved by using such an optimal pulse as a GRAPE pulse to avoid populating leakage levels or suppressing the phase acquired by the AC Stark shift~\cite{Motzoi2009-ju}. However, those are beyond the scope of this paper.

\section{\label{sec:detection}Read-out of the qubit state}

By spin-selectively exciting the Rydberg state and detecting the image-charge change induced by the Rydberg excitation using an LC circuit, a QND read-out for the spin state can be realized.

The vertical position of the electron is changed by $z_{22}-z_{11}=11.5$~nm for $V_1=60~$mV, and $V_2=-45~$mV, when the Rydberg state is excited from the ground state to the 1st excited state (Fig.~\ref{fig:imagecharge_wavefunctions}(c,d)). By applying a higher voltage to $V_1$, the electron is more strongly attracted  to the helium surface and the change in vertical position by excitation of the Rydberg state becomes smaller. 
The excitation from the Rydberg-ground state to the Rydberg-1st-excited state changes the image charge induced on the center electrode by  $\Delta q \sim 0.01 e$ (Fig.~\ref{fig:imagecharge_wavefunctions}(c,d)) for $V_1=60~$mV and  $V_2=-45~$mV. In order to calculate the induced image charge on the center electrode, we applied the Shockley-Ramo theorem~\cite{He2001}. The theorem states that the charge $Q$ on an electrode induced by a charge $q$ is given by $Q=-q\phi_0 (x)$, where $\phi_0 (x)$ is the electric potential that would exist at the  position $x$ of charge $q$ under the following circumstances: the selected electrode at unit potential, all other electrodes at zero potential and all charges are removed. Following this theorem, we run COMSOL simulation with  $1$~V to the center electrode and all the other electrodes grounded and then obtained the electric potential $V_e$ at the position of the electron. The induced image charge on the center electrode by the electron is calculated to be $\frac{V_e}{1V}e$. 

 By setting a MW frequency to $f_{0,\mathrm{Ry}A}^{n_z^B=1,\sigma^A=\uparrow}$, the Rydberg excitation and the image charge change occur only when qubit A is in spin-up state (Fig.~\ref{fig:energy_diagram_and_pulses}(a)). Under MW excitation, the induced image charge on the center electrode is $\Delta Q= \rho_{22} \Delta q$, where $\rho_{22}$ is the probability of the electron to occupy the Rydberg-1st-excited state of qubit A, which depends on the MW frequency detuning from the resonance: $\Delta f_0 = f_\mathrm{MW}- f_{0,\mathrm{Ry}A}^{n_z^B=1,\sigma^A=\uparrow}$(Fig.~\ref{fig:energy_diagram_and_pulses}(a)).  Furthermore, we apply a small and slow voltage modulation (probe signal) $u_m \cos(2 \pi f_mt)$ to the center electrode. The detuning from the resonance is modulated due to the Stark shift of the Rydberg energy levels and the time-averaged  induced image charge on the center electrode can be approximated by $\Delta Q= \overline{\rho_{22} } (t)\Delta q$. Here, $\overline{\cdot}$ denotes the average over time much longer than $1/f_\mathrm{MW}$ but much shorter than $1/f_m$. Therefore, the charge on the center electrode is expressed as $Q=C_0 u_m\cos(2 \pi f_m t) +\Delta Q$ where $C_0$ is the stray capacitance of the center electrode. The stray capacitance of such a nano-fabricated device is dominated by  the bond pads and is typically $\sim 1~$pF~\cite{Gonzalez-Zalba2015}. Focusing on the in-phase component of $\Delta Q$ in response to the voltage modulation, the charge on the center electrode is rewritten as $Q \approx (C_0 + \Delta C) u_m \cos(2 \pi f_mt)$ with the effective capacitance change:
\begin{equation}
\Delta C\approx \frac{\left<\rho_{22} \right>_I \Delta q}{u_m} ,
 \end{equation}
where $\left<\rho_{22} \right>_I= \int_{-t_c}^{t_c} \rho_{22} (t) \cos(2 \pi f_mt)  dt /t_c$ is the in-phase component of $\rho_{22}(t)$ and $t_c \gg 1/f_m$. 
This effective capacitance change happens only when qubit A is in the spin-up state.

In order to estimate the effective capacitance change, we calculate the lever arm of the voltage applied to the center electrode to the Rydberg transition energy shift  $\alpha ' = 2$~GHz/mV. The effective capacitance change can be rewritten as  $\Delta C = \alpha'  \Delta q\frac{\left<\rho_{22} \right>_I}{A_m} $. There is an optimal set of parameters for $\frac{\left<\rho_{22} \right>_I}{A_m}$ to be maximized. We found that the optimal point occurs when $f_m = f_\mathrm{1,Ry}$, the MW frequency is slightly detuned, and the modulation amplitude $A_m = f_1 /10=5~$MHz, which gives  $\left<\rho_{22} \right>_I \approx 0.1$ and $\Delta C \approx 60$~aF (Appendix~\ref{sec:read-out_supplem}). The effective capacitance change can be detected as the change in the resonance frequency of the LC resonant circuit. The LC resonant circuit is formed by connecting an inductance to the center electrode as done in semiconductor quantum dots \cite{Gonzalez-Zalba2015,Brenning2006-wc}. Recently the capacitance sensitivity  was measured to be as high as $0.04~\mathrm{aF}/\sqrt{\mathrm{Hz}}$~\cite{Apostolidis2020-ck}. This allows us to detect   $\Delta C \approx 60$~aF at the signal-to-noise ratio $\approx 1$ with the measurement bandwidth $1~$MHz, with which we can expect to have a fast enough qubit-state read-out for topological quantum error correction~\cite{Fowler2012}. Summarizing the above, we can read out the qubit state (the spin state) by detecting the Rydberg transition via the measurement of the LC resonance frequency.  This read-out technique works as a QND measurement \cite{Braginsky1996-cr} for the spin state. A QND measurement is obtained if the Hamiltonian of the system including a measurement apparatus commutes with an observable. In our case, the observable is the $z$ projection of the spin state of qubit $i$,  while the measurement apparatus is the Rydberg state of qubit $i$. The Hamiltonian of qubit $i$ is given by 

\begin{equation}
H^i=  \frac{1}{2}\Delta R_i s_z^i +\sum_j J s_z^i s_z^j+\frac{1}{2}g \mu_\mathrm{B} B \sigma_z^i -\frac{1}{2} \Delta b \sigma_z^i s_z^i,  \label{final_Hamiltonian}
\end{equation}
where $\sum_j$ is the sum over the nearest neighbor qubits of qubit $i$. From Eq.~(\ref{final_Hamiltonian}), we obtain $[\sigma_z^i,H^i]=0$. 

$\sum_j J s_z^i s_z^j$ in Eq.~(\ref{final_Hamiltonian}) tells us that we cannot realize the two-qubit gate or the read-out of the qubit state for the nearest neighbor qubits at the same time. If one or more nearest neighbors of an electron are in the Rydberg-excited state, then also exciting this electron will result in the electrons' states being altered unintentionally. Although this fact sets some constraints on how to realize quantum gates, we expect only a minor impact since the read-out of the qubit state and the two-qubit gate can be performed much faster than the spin coherence time.

\section{summary and discussions}

In summary, we propose a hybrid qubit system consisting of the charge and spin states of electrons on the surface of liquid helium. An artificially-introduced magnetic field gradient induces the spin-orbit interaction and the Rydberg-spin interaction, which allows us to coherently control the spin state electrically and perform a two-qubit gate for the spin state via the Coulomb interaction. The electrons are trapped on top of pillars, separated by a distance of $0.88~\mu$m, which allows the electric dipole-dipole interaction between the electrons to be as strong as $\approx$140~MHz. The introduced magnetic field gradient mixes the spin state and the orbital state and shortens the spin relaxation time to 50~ms for the configurations that we considered here. At the same time, the existence of the magnetic field gradient does not degrade the spin dephasing time and it stays around $100~$s at liquid helium temperature $\sim 100$~mK. The Rabi frequency of the spin state is more than $10^6$ times higher than the spin relaxation and we estimated the single-qubit gate fidelity to be $>99.9999\%$. However, we suspect that in real experiments  it must be limited by other experimental sources such as the phase noise of the MW generator \cite{Ball2016}. As a two-qubit gate is realized by the coherent control of the Rydberg transition, its fidelity is limited by the Rydberg relaxation rate. With the Rydberg relaxation rate 
 $ T_\mathrm{1,Ry}^{-1} \sim 1~$MHz, the fidelity of a controlled-phase gate was estimated to be $\sim 99 \%$.   Further improvements in the two-qubit gate scheme are foreseeable. The quantum-non-demolition measurement of the qubit state is achieved via the Rydberg-spin interaction by detecting the Rydberg transition of the qubit using a resonant LC-circuit with a measurement bandwidth $\approx 1~$MHz. 

We believe that the experimental demonstration for a small number of qubits can be readily done since the proposed device geometry is similar to the devices with which trapping a single electron on helium was achieved~\cite{Glasson2005-sw,Papageorgiou2005,Rousseau2007,Koolstra2019-mq} and we can make use of the well-established fabrication technique of a nano-scale ferromagnet for semiconductor quantum dots~\cite{Pioro-Ladriere2008,Petersen2013-ti,Lachance-Quirion2015-yz}. Together with the feasibility of employing the three-dimensional wiring techniques under development~\cite{Veldhorst2017,Yost2020-xv,Tamate2022-ba}, this architecture makes electrons on helium a strong candidate to realize a fault-tolerant quantum computer.

\section*{Acknowledgement}
We acknowledge Yury Mukharsky, John Morton, Yuichi Nagayama, Asher Jennings, Kazuya Ando, and Atsushi Noguchi for useful discussions. 

This work was supported by JST-FOREST (JPMJFR2039), RIKEN-Hakubi program, and “KICKS” grant from Okinawa Institute of Science and Technology (OIST) Graduate University. 
\newpage
\appendix

\section{Wavefunction and energy levels}\label{wavefunction_energy}

\begin{figure}[H]
\centering
\includegraphics[width=0.8\textwidth]{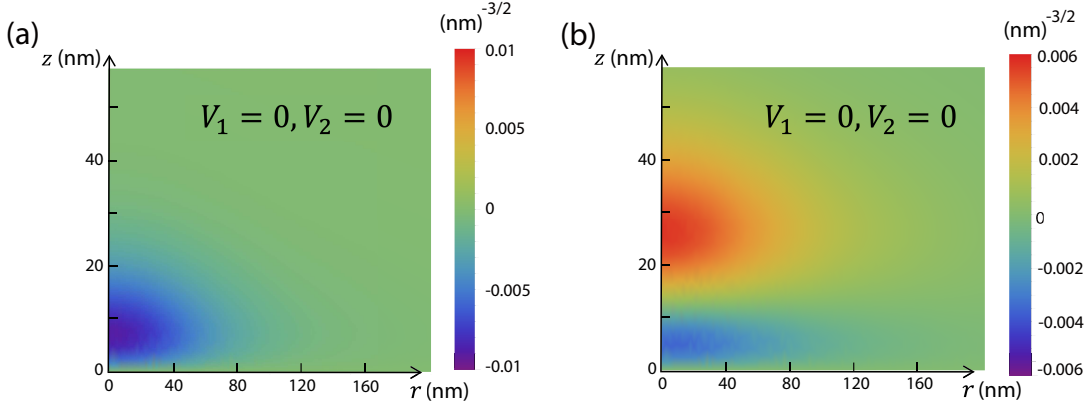}
\caption{\label{fig:wavefunctions}  Normalized eigenfunction $\phi(r,z)$ for $(m,n_r,n_z)=(0,0,1)$ in (a) and $=(0,0,2)$ in (b) with $V_1=V_2=0$ and $B=0$. }
\end{figure}

\begin{figure}[H]
\centering
\includegraphics[width=0.8\textwidth]{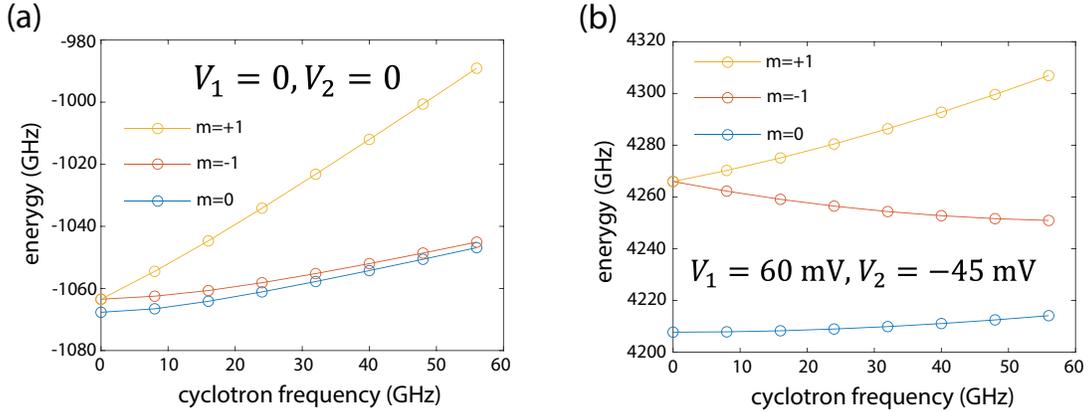}
\caption{\label{fig:energy_levels_m0_pm1_cyclotron}  (a) The energy eigenvalues for $m=0, \pm1$ and $(n_r,n_z)=(0,1)$ as a function of the cyclotron frequency $\omega_c=eB_0/m_e$ with $V_1=V_2=0$ . (b) The same as in (a) with $V_1=60$ mV, $V_2=-45$ mV.}
\end{figure}

Fig.~\ref{fig:wavefunctions}(a,b) show the normalized eigenfunctions  $\phi(r,z)$ for $(m,n_r,n_z)=(0,0,1)$ and $(m,n_r,n_z)=(0,0,2)$, respectively, for $V_1=V_2=0$ and $B=0$. The electron is locally confined due to the image potential even if the electrode potentials are set to ground. The eigenenergies as a function of the cyclotron frequency determined by the applied magnetic field $B$ are shown in Fig.~\ref{fig:energy_levels_m0_pm1_cyclotron}.

\begin{figure}
\centering
\includegraphics[width=1\textwidth]{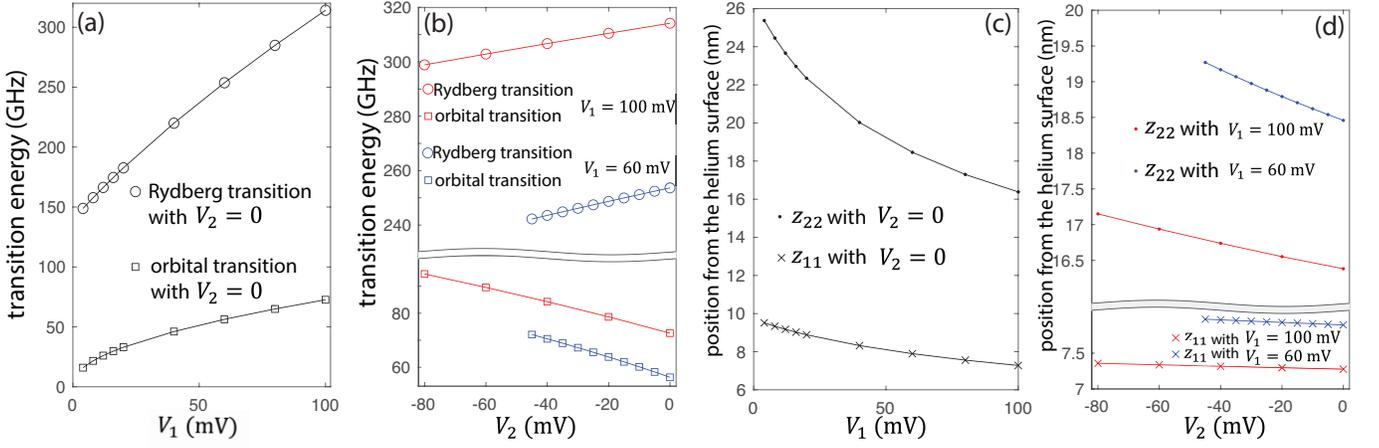}
\caption{\label{fig:energy_transition_vs_voltages}  The magnetic field is set to 0 in (a-d). (a) Transition energies as a function of $V_1$  with $V_2=0$. Circles represent the transition energy between the two lowest Rydberg states, $E^{(0)}_{0,0,2}- E^{(0)}_{0,0,1} $. Squares represent the transition energy between the two lowest in-plane orbital states,  $E^{(0)}_{0,1,1}- E^{(0)}_{0,0,1} $. (b) Red and blue circles represent the transition energies between the two lowest Rydberg states as a function of $V_2$  with $V_1=100~$mV and with $V_2=60~$mV, respectively. Red and blue squares  represent the transition energies between the two lowest orbital states as a function of $V_2$  with $V_1=100~$mV and with $V_2=60~$mV, respectively. (c) The dots represent the average position of the Rydberg-1st-excited state along the $z$ axis and the crosses represent that of the Rydberg-ground state as a function of $V_1$ with $V_2=0$. (d) The red and blue dots represent  the average position of the Rydberg-1st-excited state along the $z$ axis as a function of $V_2$ with $V_1=100~$mV and $=60~$mV, respectively. The red and blue crosses represent  the average position of the Rydberg-ground state along the $z$ axis as a function of $V_2$ with $V_1=100~$mV and $=60~$mV, respectively.}
\end{figure}

Fig.~\ref{fig:energy_transition_vs_voltages}(a) shows how the Rydberg and orbital transition energies change as a function of $V_1$, when $V_2$ is set to 0. With increasing $V_1$, both the Rydberg transition energy and the orbital transition energy become higher (both the confinement along the $z$ axis and the confinement along the $x, y$ axes become stronger). Fig.~\ref{fig:energy_transition_vs_voltages}(b)  shows the  transition energies as a function of $V_2$ for $V_1=60~$mV and $100~$mV. In this case, the Rydberg transition energy decreases as $V_2$ decreases, while the orbital transition energy increases.  Fig.~\ref{fig:energy_transition_vs_voltages}(c) and (d) show the average vertical position of the electron wavefunction as a function of $V_1$ and $V_2$, respectively. The electron is confined more tightly by increasing $V_1$ and $V_2$ and comes closer to the liquid surface. 

\section{\label{sec:stray_magnetic_field}Stray magnetic fields}

\begin{figure*}[h!]
\centering
\includegraphics[width=0.9\textwidth]{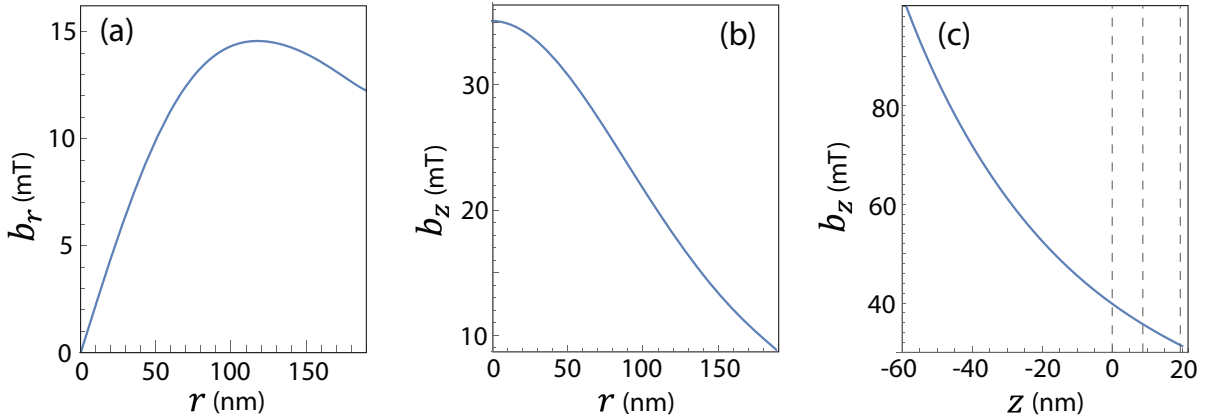}
\caption{\label{fig:magnetic_field_gradients}
The $r$ component of the stray magnetic field  ($b_r$) and the $z$ component of the stray magnetic field ($b_z$) created by the ferromagnetic pillar as a function of $r$ at $z=z_{11}$ in (a,b) and as a function of $z$ at $r=0$ in (c) for the saturation magnetization of Co (1.8~T). (c) The dashed lines indicate the liquid helium surface ($z=0$), the average position of the electron in the Rydberg-ground state ($z=z_{11}$), and in the Rydberg-1st excited state ($z=z_{22}$). }
\end{figure*}

Fig.~\ref{fig:magnetic_field_gradients} (a) and (b) show the vertical stray magnetic field $b_r$ and the in-plane stray magnetic fields $b_z$, respectively, which are introduced in Sec,~\ref{sec:hamiltonian}, as a function of $r$ at $z=z_{11}$. Fig.~\ref{fig:magnetic_field_gradients} (c) shows $b_z$ as a function of $z$ at $r=0$.

\section{Single-qubit gates}

\subsection{Spin rotation}\label{sec:coherent_control}

As shown in Sec.~\ref{sec:single-qubits}, single-qubit gates can be realized by creating an AC electric field at the electron's position. Following Refs.~\cite{Winkler2003,Romhanyi2015},  time-dependent Schrieffer-Wolff transformation gives us an effective time-dependent Hamiltonian for the subspace spanned by $i=\{m=0, \downarrow\}$ and $f=\{m=0,\uparrow \}$ when the subspace is well separated in energy from the rest of the Hilbert spaces: $E_f-E_i \ll E_\beta-E_{i,f}$, here $\beta \neq i,f$. By taking $H(t)$ in Eq.~\ref{eq:Ht_onequbit} as a perturbation part and $H_0$ in Eq.~\ref{Hamiltonian_0} as an unperturbed part, the effective Hamiltonian can be calculated as $H_\mathrm{eff}=H_0+\frac{1}{2}[S,H(t)]$ with $[S,H_0]=-H(t)$. Consequently, the Rabi frequency of the spin state is calculated to be $f_\mathrm{1,spin}=\frac{\gamma}{2 \pi \hbar} $ where $\gamma$ is given by~\cite{Winkler2003,Romhanyi2015} 
\begin{align}
    \bra{i} H_\mathrm{eff}\ket{f} &=\frac{1}{2}   \sum_{\beta \neq i,f} (H(t))_{ i \beta} (H(t))_{ \beta f}
\left( \frac{1}{E_i-E_{\beta}} + \frac{1}{E_f-E_{\beta}} \right)\\
&\approx   \sum_{\beta = \beta_1,\beta_2} ((U)_{ i \beta} (W)_{ \beta f} +(W)_{ i \beta} (U)_{ \beta f})
\left( \frac{1}{E_i-E_{\beta}} + \frac{1}{E_f-E_{\beta}} \right)\cos (2 \pi f_\mathrm{0, spin} t)\\
&\equiv 2 \gamma \cos (2 \pi f_\mathrm{0, spin} t)
,
\end{align}
where $(H(t))_{\triangle, \bigcirc}$  stands for $\bra{\triangle} H(t)\ket{\bigcirc}$ and $\sum_\beta$ refers to the sum over all the possible virtual states. Through this approach, the problem is reduced to the conventional two-level Rabi model. 

\begin{figure}[H]
\centering
\includegraphics[width=0.7\textwidth]{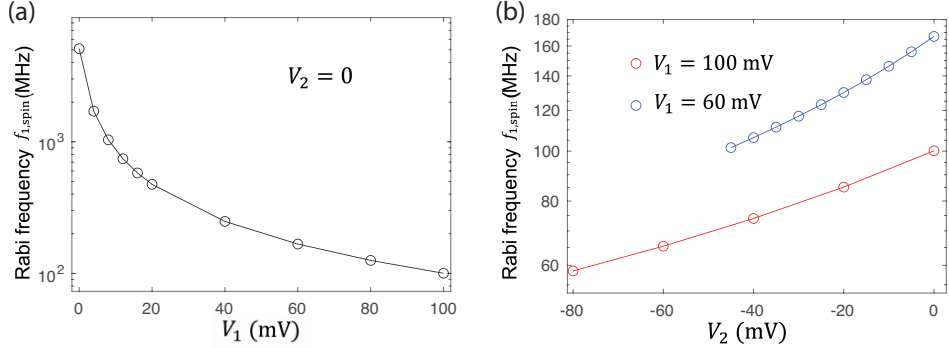} 
\caption{\label{fig:Rabi_vs_voltage}
Rabi frequency of the spin state as a function of $V_1$ with $V_2=0$ in (a) and as a function of $V_2$ with $V_1=100$~mV (red circles) and with $V_1=60$~mV (blue circles) in (b) when the cyclotron frequency of the electron corresponds to  $\omega_c=eB_0/m_e=16~$GHz.}
\end{figure}

Fig.~\ref{fig:Rabi_vs_voltage} shows the numerically calculated Rabi frequency of the spin state for various $V_1$ and $V_2$, assuming  $f_\mathrm{0,spin} \approx \omega_c/ 2 \pi =16$~GHz and $E^\mathrm{RF}=500$~V/cm. We simulated the Rabi frequency of the spin state for the magnetic field $B$ in the range of $0$ to $2$~T and found that the Rabi frequency varies with the magnetic field by less than 20~\% except for the case when the applied voltages are $V_1=V_2=0$ (Fig.~\ref{fig:Rabi_vs_voltage}). We also found that the Rabi frequency increases when the orbital confinement gets weaker.

\subsection{Single-qubit gate fidelity}\label{sec:single_qubit_gate_fidelity}

In order to calculate the gate fidelity, we used the Lindblad master equation \cite{Lindblad1976-xi}
\begin{equation}
    \frac{d\rho}{dt}=-\frac{i}{\hbar}[H, \rho]+\frac{1}{2}\sum_i L(A_i)\rho \label{Lindblad_eq}
\end{equation}
with $L(\Lambda)\rho =2 \Lambda \rho \Lambda^ \dagger -  \Lambda^\dagger \Lambda \rho
-
\rho \Lambda^\dagger \Lambda $. To calculate the single-qubit gate fidelity, we take the spin-state density operators as $\rho$, and we take
$A_1= \sqrt{T_\mathrm{1,spin}^{-1}}\ket{\downarrow}\bra{\uparrow}$ and $A_2=\sqrt{\frac{T_\mathrm{2,spin}^{-1}}{2}} \sigma_z$ as collapse operators. The Hamiltonian is described as  $H=2 \pi f_\mathrm{1,spin} \frac{\cos(\Theta) \sigma_x + \sin(\Theta) \sigma_y}{2}$ and $\Theta$ can be tuned by  the AC electric field phase depending on the gates that we aim to apply.  Here, we consider the case for the Rabi frequency $f_\mathrm{1,spin}= 100~$MHz, the relaxation time:  $T_\mathrm{1,spin} = 50~$ms (Sec.~\ref{sec:spin_relaxation}), and the dephasing time  $T_\mathrm{2,spin}=100~$s (Sec.~\ref{sec:spin-dephasing}). The average single-qubit gate fidelity is calculated to be 
\begin{equation}
F_1=\frac{1}{2} \sum_{i=0,1} \frac{1}{24} \sum_{j=1}^{24} \mathrm{Tr} [\mathcal{C}_j^\mathrm{ideal}(\rho_i)\mathcal{C}_j^\mathrm{real}(\rho_i)], 
\end{equation}where $\mathcal{C}_j^\mathrm{ideal} (\rho_i)$ and $\mathcal{C}_j^\mathrm{real} (\rho_i)$ denote the density operators obtained after applying one of the 24 Clifford gates in the single-qubit Clifford groups~\cite{Knill2008-rb,Emerson2005-zg} to the initial state $\rho_i$ in the ideal case and real cases, respectively, and we take the average of two initial states $\rho_0=\ket{\downarrow}\bra{\downarrow}$ and $\rho_1=\ket{\uparrow}\bra{\uparrow}$~\cite{Muhonen2015-yl}. $\mathcal{C}_j^\mathrm{real}(\rho_i)$ was numerically simulated using qutip.mesolve~\cite{Johansson2013-rq} and we obtained $F_1>99.9999\%$. The average rotation of the 24 Clifford gates is $26 \pi/24$ and thus their average gate time is 5.4~ns.

\section{Longitudinal relaxation of the spin state} \label{sec:longitudinal_relaxation_appendix}

The spin relaxation is an energy non-conserving process and the energy difference between the final state and the initial state (the Zeeman energy) is transferred to the emitted two ripplons. The Hamiltonian of the electron-ripplon interaction, which corresponds to the two-ripplon process is given by $H_{\mathrm{2ri}}$ (Eq.~(\ref{eq_H2ri})). It was previously demonstrated that the two-ripplon emission of short wavelength capillary waves is responsible for a process that requires an energy change of  $\gg k_B T$ ~\cite{Monarkha2007-el}. Here, $k_B$ is the Boltzmann constant and $T$ is the liquid helium temperature.  The electron-two-ripplon coupling for short-wavelength capillary waves is given by $U_{\bm{q_1}\bm{q_2}}=\frac{1}{2}\hat{F} $, where $\bra{m,n_r,n_z} \hat{F} \ket{m',n_r',n_z'}=2 \kappa_0 \sqrt{ \bra{m,n_r,n_z} \left( \frac{\partial v}{\partial z}\right) \ket{m,n_r,n_z} \bra{m',n_r',n_z'}\left( \frac{\partial v}{\partial z}\right)\ket{m',n_r',n_z'} }$, $\kappa_0=\sqrt{2 m_e V_0}/\hbar$, $V_0\approx$1~eV is the barrier at the helium surface and $v$ is the energy of the electron in the electric potential created by the image charge and externally applied electric field~\cite{Monarkha2007-el} . 

Accordingly, the Hamiltonian corresponding to the two-ripplon emission of short wavelength capillary waves is written as
\begin{equation}
H_{\mathrm{2ri}}^F=\frac{1}{2}\sum_{\bm{q_1}} \sum_{\bm{q_2}} Q_{\bm{q_1}} Q_{\bm{q_2}}  \hat{F}
e^{i (\bm{q_1}+\bm{q_2})\bm{r}} a_{-\bm{q_1}}^\dagger a_{-\bm{q_2}}^\dagger.
\end{equation}

\begin{figure}[H]
\centering
\includegraphics[width=0.7\textwidth]{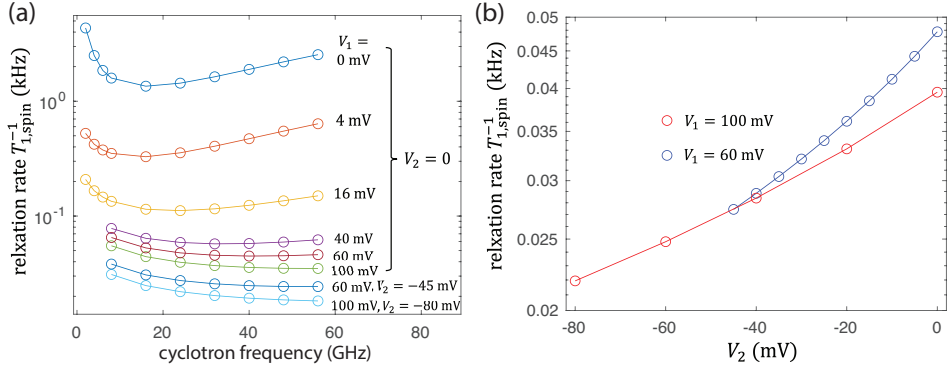} 
\caption{\label{fig:relaxation_vs_voltage_cyclotron}
Relaxation rate of the spin state as a function of $\omega_c$ for different $V_1$ and $V_2$ in (a) and as a function of $V_2$ with $V_1=100$~mV (red circles), with $V_1=60$~mV (blue circles) and $\omega_c=16~$GHz in (b).
\label{fig:relaxation_vs_cyclotron_voltage}}
\end{figure}

In order to satisfy the conservation of the total momentum of an electron and ripplons, two ripplons must be emitted in nearly opposite directions, that is $\bm{q_1} \sim -\bm{q_2}$. The total Hamiltonian which is responsible for the relaxation of the spin state can be written as 
\begin{equation}
H=H_{\mathrm{2ri}}^F+W_{rr}
\label{relaxation_Hamiltonian}.
\end{equation}
As shown in Eq.~(\ref{eq_Gamma_relax2}), the spin relaxation rate is

\begin{equation}
T_\mathrm{1,spin}^{-1}\approx \frac{2\pi}{\hbar} \left|  
\frac{(W_{rr})_{i \beta_1 } (H_\mathrm{2ri}^F)_{\beta_1 f }   }{E_i-E_{\beta_1}}+
\frac{ (H_{\mathrm{2ri}}^F)_{i \beta_2} (W_{rr})_{ \beta_2 f} }{E_i-E_{\beta_2}+\epsilon_{n_\mathrm{ri}}-\epsilon_{n_\mathrm{ri}'}}
\right|^2\delta(E_f-E_i+\epsilon_{n_\mathrm{ri}'}-\epsilon_{n_\mathrm{ri}})
\label{eq_Gamma_relaxation},
\end{equation}
where the definition of the states $i$, $f$, $\beta_1$ and $\beta_2$ is the same as in Sec.~\ref{sec:spin_relaxation}.
Using the explicit form: $(W_{rr})_{\beta_2 f}=(W_{rr})_{i \beta_1}=\frac{h}{2} \partial_r f_{b_r} l_0 $ and the relation $(H_\mathrm{2ri}^F)_{i \beta_2}=(H_\mathrm{2ri}^F)_{\beta_1 f}$ , Eq.~(\ref{eq_Gamma_relaxation}) is further reduced to  
\begin{equation}
   T_\mathrm{1,spin}^{-1}\approx \frac{2\pi}{\hbar} \left|  
(H_\mathrm{2ri}^F)_{\beta_1 f}
\right|^2 \frac{1}{4} (\partial_r f_{b_r})^2 l_0^2   \left|
\frac{1}{f_i-f_{\beta_1}}+\frac{1}{f_i-f_
{\beta_2}+f_\mathrm{0,spin}}\right|^2 \delta(E_f-E_i+\epsilon_{n_\mathrm{ri}'}-\epsilon_{n_\mathrm{ri}}). 
\end{equation}
Using the explicit form of $H_{2\mathrm{ri}}^F$, we obtain

\begin{align}
&|(H_{2\mathrm{ri}}^F)_{\beta_1 f }|^2\delta(E_f-E_i+\epsilon_{n_\mathrm{ri}'}-\epsilon_{n_\mathrm{ri}})\nonumber \\
& \approx 2 \sum_{\bm{q}} \sum_{\bm{s}} Q_{\bm{q}}^4 \kappa_0^2 \sum_k i^k|\bra{\psi_{m=0}(\theta)} e^{i k \theta} \ket{\psi_{m=\pm 1}(\theta)}\bra{\phi_{0, 0, 1} }\frac{\partial v}{\partial z}
  J_k(sr) \ket{\phi_{\pm 1,0,1}}|^2 \delta( h f_\mathrm{0,spin}-2\hbar \omega_q),
\end{align}where $s=|\bm{q_1}+\bm{q_2}|$ and $J_1$ is the Bessel function of the 1st kind. By inserting the identity operator $\bm{I}=\sum_{n_r,n_z} \ket{\phi_{m=0,n_r,n_z}} \bra{\phi_{m=0,n_r,n_z}}$, $\bra{\phi_{0,0,1}}\frac{\partial v}{\partial z}
  J_1(sr) \ket{\phi_{\pm 1,0,1}}$ 
  can be approximated by 
$\bra{\phi_{0,0,1}}\frac{\partial v}{\partial z}\ket{\phi_{0 ,0,1}}\bra{\phi_{0,0,1}}  J_1(sr) \ket{\phi_{\pm 1,0,1}}$. We numerically confirmed that the other terms are negligibly small. This results in
\begin{align}
T_\mathrm{1,spin}^{-1}\approx & 
\frac{m_e V_0    (\partial_r f_{b_r})^2l_0^2}{12\hbar h\rho_\mathrm{He}^{2/3} \alpha_\mathrm{He}^{4/3} \omega_{q^*}^{1/3}} 
|\bra{\phi_{0,0,1}}\frac{\partial v}{\partial z}\ket{\phi_{0,0,1}}|^2 \nonumber \\
\times &\int s | \bra{\phi_{0,0,1}}  J_1(sr) \ket{\phi_{ \pm 1,0,1}}|^2 ds \left|\frac{1}{f_i-f_{\beta_2}-f_\mathrm{0,spin}}+
\frac{1}{f_i-f_{\beta_1}}\right|^2,
\end{align}where $\omega_{q^*}=\pi f_\mathrm{0,spin} \approx \omega_c/2$. The density and the surface tension of liquid helium 4 are $\rho_\mathrm{He}=0.145~$g/cm and $\alpha_\mathrm{He}=0.378~$erg/cm$^2$, respectively. For the Fock-Darwin model, where the magnetic field is applied along $z$ and the $x-y$ confinement is described by a harmonic potential, we have the relation $ \int s|\bra{\phi_{0,0,1}}  J_1(sr) \ket{\phi_{\pm 1,0,1}}|^2 ds =1/2l_0^2$ \cite{Jacak1998}.
Although we cannot naively assume that the $x-y$ confinement can be approximated by a harmonic potential, we confirmed that the above relation is approximately valid by numerical simulation for all the configurations of the voltages and the magnetic fields shown in Fig.~\ref{fig:relaxation_vs_voltage_cyclotron}. Hence, the spin relaxation rate is further simplified as 
\begin{align}
T_\mathrm{1,spin}^{-1} \approx
\frac{m_e V_0    (\partial_r f_{b_r})^2}{24\hbar h\rho_\mathrm{He}^{2/3} \alpha_\mathrm{He}^{4/3} \omega_{q^*}^{1/3}} |\bra{\phi_{0,0,1}}\frac{\partial v}{\partial z}\ket{\phi_{0,0,1}}|^2\left|\frac{1}{f_i-f_{\beta_2}-f_\mathrm{0,spin}}+
\frac{1}{f_i-f_{\beta_1}}\right|^2.
\end{align}
Our numerical simulation shows that $\frac{\partial  b_r}{\partial r}=0.23$~mT/nm at the position of the electron (Fig.~\ref{fig:magnetic_field_gradients}). Using this value, the relaxation rate is calculated as shown in Fig.~\ref{fig:relaxation_vs_voltage_cyclotron}. Overall, the relaxation rate becomes smaller as the electric orbital confinement gets stronger (Fig.~\ref{fig:relaxation_vs_voltage_cyclotron}). When  $V_1=100~$mV and $V_2=-80~$mV, the relaxation time becomes as long as 100~ms. The magnetic field dependence shows the trade-off between the ${\omega_q^*}^{-1/3}$ term and the 
$\left|\frac{1}{f_i-f_{\beta_2}-f_\mathrm{0,spin}}+
\frac{1}{f_i-f_{\beta_1}}\right|$ term. The ${\omega_q^*}^{-1/3}$ term becomes smaller, while the 
$\left|\frac{1}{f_i-f_{\beta_2}-f_\mathrm{0,spin}}+
\frac{1}{f_i-f_{\beta_1}}\right|$ term becomes larger, as the magnetic field is increased.

\section{Dephasing of the spin state}
\label{sec:spindephasing_appendix}

The spin dephasing  happens due to the time fluctuation of the Zeeman energy:
\begin{equation}
\delta E(t)= \delta E_\uparrow(t)-\delta E_\downarrow(t),
\end{equation}where $\delta E_\sigma (t)=E_\sigma (t)-\overline{E_\sigma (t)}$, $E_\sigma (t)$ is the energy level at a time $t$ and $\overline{E_\sigma (t)}$ is the time-averaged energy level of the spin state $\sigma$ with $\sigma=\uparrow$ or $\downarrow$.  The phase difference between the spin-up and the spin-down states evolves over time as $\varphi(t)-\varphi(0)=\frac{1}{\hbar} \int^t_0 \delta E(t') dt' $. The ensemble average of the phase evolution is given by

\begin{equation}
\left<  [\varphi(t)-\varphi(0)]^2 \right>=\frac{1}{\hbar^2} \int^t_0 dt' \int^t_0 dt''\left< \delta E(t')\delta E(t'')  \right>.
\end{equation}We assume that the successive samples of the energy splitting are uncorrelated and thus its auto-correlation function can be described as  $\left< \delta E(t')\delta E(t'')  \right>=\hbar^2  \delta(t'-t'')/T_\mathrm{2,spin}$, where $T_\mathrm{2,spin}$ is the dephasing time. Consequently, the dephasing rate $T_\mathrm{2,spin}^{-1}$  can be calculated as 

\begin{equation}
T_\mathrm{2,spin}^{-1}=\frac{1}{t \hbar^2}\int^t_0 dt' \int^t_0 dt''\left< \delta E(t')\delta E(t'')  \right>=\frac{1}{t \hbar^2}\int^t_0 dt' \int^t_0 dt''
\frac{\sum_{n_\mathrm{ri}}  \bra{n_\mathrm{ri}}   \delta E(t')\delta E(t'')   \ket{n_\mathrm{ri}} e^{-\epsilon_{n_\mathrm{ri}}/T} }{ \sum_{n_\mathrm{ri}} e^{-\epsilon_{n_\mathrm{ri}}/T}   } , \label{D_varphi}
\end{equation}where the ensemble average is taken over all the possible ripplon states and $T$ is the liquid helium temperature.

\begin{figure}[H]
\centering
\includegraphics[width=1\textwidth]{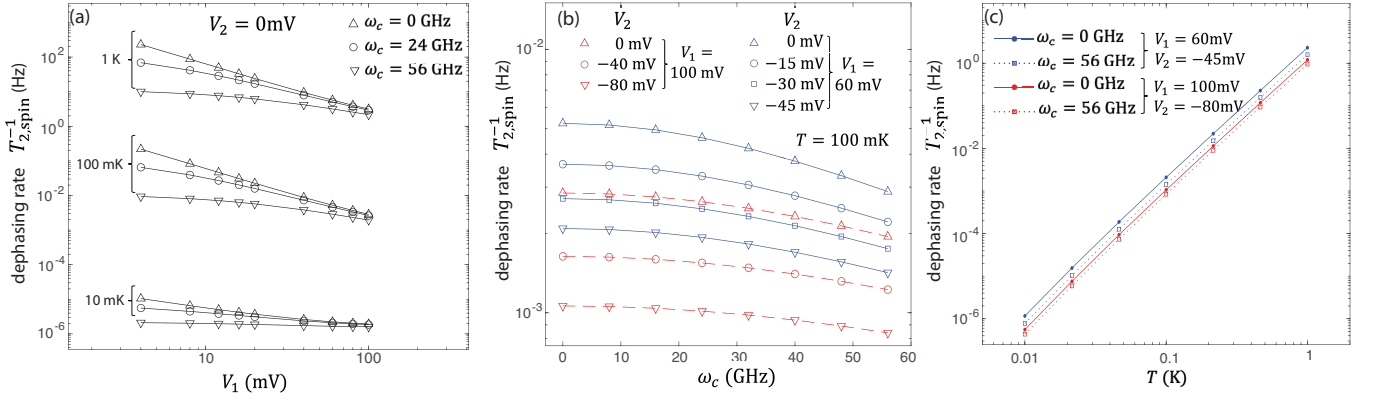} 
\caption{\label{fig:dephasing_vs_voltages} Dephasing rate of the spin state as a function of $V_1$ with $V_2=0$ for different liquid helium temperatures and magnetic fields in (a),  as a function of magnetic field (cyclotron frequency) for different voltage configurations in (b) and as a function of temperature for different voltage configurations and magnetic fields in (c). As the dephasing process is dominated by the thermally excited ripplons, it has a strong temperature dependence shown in (c). The logarithm of the dephasing rate $T_\mathrm{2,spin}^{-1}$ shows an almost linear relationship to the logarithm of the liquid helium temperature $T$.}
\end{figure}

Differently from the relaxation process, the time-averaged energy consumed or produced by the electron after the whole dephasing process is zero. Thereby, the thermally excited long-wavelength ripplons are involved here. The one-ripplon scattering cannot induce this energy-conserving process since it requires zero-energy consumption. Therefore, the two-ripplon scattering is responsible for this process. Under the long-wavelength approximation, the electron-two-ripplon coupling  is given by $U_{\bm{q_1}\bm{q_2}}=-\bm{q_1}\bm{q_2} \hat{K_z} $, where $\hat{K_z}=\frac{\hbar^2}{2 m_e}\partial^2_z$ \cite{Dykman2003}. Thus the corresponding Hamiltonian for the two-ripplon scattering is written as~\cite{Dykman2003}
\begin{equation}
H_{\mathrm{2ri}}^K=\sum_{\bm{q_1}} \sum_{\bm{q_2}} Q_{\bm{q_1}} Q_{\bm{q_2}} (-\hat{K_z})
e^{i (\bm{q_1}+\bm{q_2} )\bm{r}} (a_{\bm{q_1}}+ a_{\bm{-q_1}}^\dagger)(a_{\bm{q_2}}+ a_{\bm{-q_2}}^\dagger), 
\end{equation}
 The Hamiltonian which causes the energy fluctuation can be described as $H=H_{\mathrm{2ri}}^K+W_{rz}$. The leading term in the second-order energy shift of the spin-up (spin-down) state is calculated as 

\begin{equation}
\delta E_\sigma (t) \approx 2\frac{(H_{\mathrm{2ri}}^K)_{i\beta} (W_{rz})_{\beta i} }{E_i-E_\beta}.
\end{equation}
The definition of the states $i$, and $\beta$ here is  $i=\{n_r=0, \sigma\}$ and $\beta=\{n_r=1,\sigma\}$, where $\sigma=\uparrow$ or $\downarrow$.  The first (second) term of Eq.~(\ref{eq_delta_E_2}) of 
Sec.~\ref{sec:spin-dephasing} corresponds to $\delta E_\sigma (t)$ for $\sigma=\uparrow$ ($\downarrow$). Thus, more explicitly, the time-dependent Zeeman energy fluctuation becomes
\begin{equation}
\delta E(t) \approx \frac{\bra{\phi_{0,0,1}} H_{\mathrm{2ri}}^K \ket{ \phi_{0,1,1}} g\mu_B \frac{\partial ^2 b_z}{\partial r^2}\bra{\phi_{0,1,1}} r^2 \ket{\phi_{0,0,1}}}{E_{0,0,1}^{(0)}-E_{0,1,1}^{(0)}}.
\end{equation}By inserting the identity operator $\bm{I}=\sum_{n_r,n_z} \ket{\phi_{m=0,n_r,n_z}} \bra{\phi_{m=0,n_r,n_z}}$, we obtain the following approximation:
\begin{align}
\bra{\phi_{0,0,1}} &H_{\mathrm{2ri}}^K \ket{ \phi_{0,1,1}} \nonumber\\
\approx  &   \bra{\phi_{0,0,1}} -\hat{K_z} \ket{ \phi_{0,  0,1}} \sum_{\bm{q_1}} \sum_{\bm{q_2}} \bm{q_1}\bm{q_2}Q_{\bm{q_1}} Q_{\bm{q_2}}
\bra{\phi_{0,0,1}} e^{i(\bm{q_1}+\bm{q_2} )\bm{r}}\ket{ \phi_{0,1,1}} 
 (a_{\bm{q_1}}+ a_{\bm{-q_1}}^\dagger)(a_{\bm{q_2}}+ a_{\bm{-q_2}}^\dagger).
\end{align}We numerically confirmed that the other terms are negligibly small. 
Using this approximation, we obtain the energy correlation function as 
\begin{align}
\delta E(t') \delta E(t'')  \approx &\frac{(g\mu_B \frac{\partial ^2 b_z}{\partial r^2}\bra{\phi_{0,1,1}} r^2 \ket{\phi_{0, 0,1}})^2}{ (E_{0,0,1}^{(0)}-E_{0,1,1}^{(0)})^2}  \left| \bra{\phi_{0,  0,1}} -\hat{K_z} \ket{\phi_{0,0,1}}  \right|^2  \nonumber\\
& 
\times \sum_{\bm{q_1},\bm{q_2}} \sum_{ \tilde{\bm{q}}_1,\tilde{\bm{q}}_2} \bm{q_1}\bm{q_2} 
\tilde{\bm{q}}_1,\tilde{\bm{q}}_2
 Q_{\bm{q_1}} Q_{\bm{q_2}}Q_{\tilde{\bm{q}}_1} Q_{\tilde{\bm{q}}_2} \nonumber\\
 &
\times \bra{\phi_{0,0,1}} e^{i (\bm{q_1}+\bm{q_2} )\bm{r}}\ket{ \phi_{0,1,1}} 
\bra{\phi_{0,0,1}} e^{i (\tilde{\bm{q}}_1+\tilde{\bm{q}}_2 )\bm{r}}\ket{ \phi_{0,1,1}} \nonumber\\
& \times  (a_{\bm{q_1}}(t')+ a_{\bm{-q_1}}^\dagger(t'))(a_{\bm{q_2}}(t')+ a_{\bm{-q_2}}^\dagger (t'))(a_{\bm{q_1}}(t'')+ a_{\bm{-q_1}}^\dagger(t''))(a_{\bm{q_2}}(t'')+ a_{\bm{-q_2}}^\dagger (t'')),
\end{align}where $a_{\bm{q}}(t)=a_{\bm{q}}(0) e^{-i \omega_q t}$. Plugging this into Eq.~(\ref{D_varphi}) and taking the integral according to,

\begin{align}
&\int^t_0 dt' \int^t_0 dt''
\frac{\sum_{n_\mathrm{ri}}  \bra{n_\mathrm{ri}}   (a_{\bm{q_1}}(t')+ a_{\bm{-q_1}}^\dagger(t'))(a_{\bm{q_2}}(t')+ a_{\bm{-q_2}}^\dagger (t'))(a_{\bm{q_1}}(t'')+ a_{\bm{-q_1}}^\dagger(t''))(a_{\bm{q_2}}(t'')+ a_{\bm{-q_2}}^\dagger (t'')) \ket{n_\mathrm{ri}} e^{-\epsilon_{n_\mathrm{ri}}/T} }{ \sum_{n_\mathrm{ri}} e^{-\epsilon_{n_\mathrm{ri}}/T}   } \nonumber \\
&= t \pi (\bar{n}_{q_1}+1)\bar{n}_{q_2} \delta(\omega_{q_1}-\omega_{q_2}) \delta(\bm{q_1}-\tilde{\bm{q}}_1)\delta(\bm{q_2}-\tilde{\bm{q}}_2),
\end{align}where $\bar{n}_q=(e^{\hbar \omega_q/T}-1)^{-1}$ and $\omega_q=\sqrt{\alpha_\mathrm{He} q^3/ \rho_\mathrm{He}}$, we obtain the dephasing rate as
\begin{align}
T_\mathrm{2,spin}^{-1} \approx &\frac{  \pi  (g\mu_B \frac{\partial ^2 b_z}{\partial r^2} l_0^2 )^2}{ \hbar^2 (E_{0,0,1}^{(0)}-E_{0,1,1}^{(0)})^2}  \left| \bra{\phi_{0,0,1}} -\hat{K_z} \ket{ \phi_{0,0,1}}  \right|^2  \nonumber\\
&
\times \sum_{\bm{q_1},\bm{q_2}} (\bm{q_1}\bm{q_2} )^2
( Q_{\bm{q_1}} Q_{\bm{q_2}})^2  \left | \frac{|\bm{q_1}+\bm{q_2}|^2 l_0^2}{4}
 e^{-|\bm{q_1}+\bm{q_2}|^2 l_0^2/4} \right|^2  
(\bar{n}_{q_1}+1)\bar{n}_{q_2} \delta(\omega_{q_1}-\omega_{q_2}) . 
\label{dephasing_time}
\end{align}Here we used the expressions $\bra{\phi_{0,1,1}} r^2 \ket{\phi_{0,0,1}}=l_0^2$ and $|\bra{\phi_{0,0,1}} e^{i(\bm{q_1}+\bm{q_2} )\bm{r}}\ket{\phi_{0,1,1}} |=\left | \frac{|\bm{q_1}+\bm{q_2}|^2 l_0^2}{4}
 e^{-|\bm{q_1}+\bm{q_2}|^2 l_0^2/4} \right|$, which are valid for a harmonic confinement and is a good approximation for all the configurations shown in Fig.~\ref{fig:dephasing_vs_voltages}. 
By taking $x=q l_0$, Eq.~(\ref{dephasing_time}) is further reduced to
\begin{align}
T_\mathrm{2,spin}^{-1} \approx \frac{\hbar^4 \left| \bra{\phi_{0,0,1}} \partial^2_{z}\ket{ \phi_{0,0,1}}  \right|^2}{16 m_e^2  48 \pi^2 \rho_\mathrm{He}^{1/2} \alpha_\mathrm{He}^{3/2} l_0^{11/2} }\frac{(\partial^2_{r} f_{b_z} l_0^2 )^2  }{|\Delta f_{n_r=1 \rightarrow  0}|^2  }
\int_0^\infty dx \int_0^{2\pi} d \theta (\bar{n}_{q}+1)\bar{n}_{q} x^{17/2} e^{-x^2(1-\cos \theta)} \cos^2 \theta (1-\cos \theta)^2,
\end{align}
where $\partial^2_{r} f_{b_z}= g\mu_B \frac{\partial ^2 b_z}{\partial r^2}/h$ and $\Delta f_{n_r=1 \rightarrow  0}=(E_{0,0,1}^{(0)}-E_{0,1,1}^{(0)})/h$. Our numerical simulation shows that $\frac{\partial ^2 b_z}{\partial r^2}=-0.0040$~mT/(nm)$^2$ at the position of the electron (Fig.~\ref{fig:magnetic_field_gradients}). Using these values, we calculated the dephasing rate as shown in Fig.~\ref{fig:dephasing_vs_voltages}.

\section{Screening effect on the Coulomb interaction\label{sec:screening}}

\begin{figure}[H]
\centering
\includegraphics[width=0.9\textwidth]{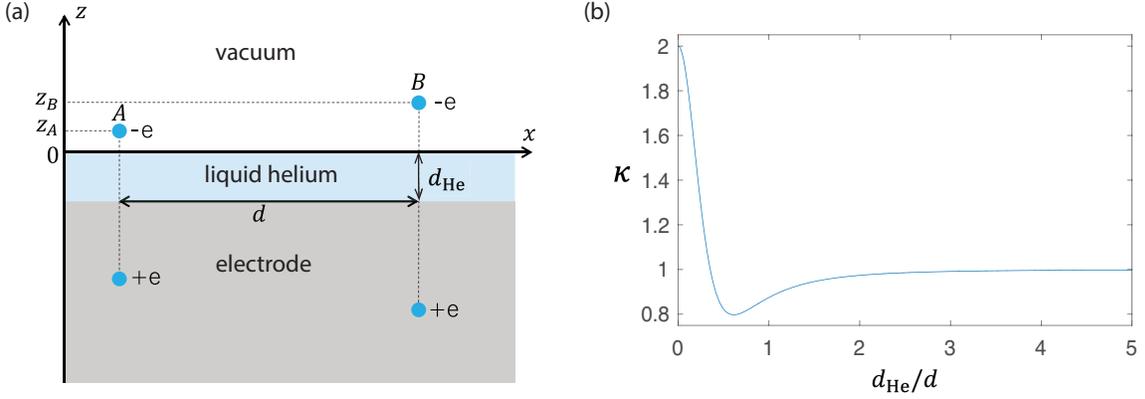}
\caption{\label{fig:screening} (a) Electron A and electron B are  separated by $d$ horizontally.  An electron and the induced image charge on the electrode are equally distant from the surface of the electrode. The value of the image charge is equal to the elementary charge. $z_A$ and $z_B$ denote the vertical positions of electrons A and B, respectively. The figure depicts the case where electron A is in the Rydberg-ground state and electron B is in the Rydberg-1st-excited state.  (b) The enhancement factor acquired by the screening effect in the electric dipole-dipole interaction as a function of the helium depth $d_\mathrm{He}$ normalized by $d$.}
\end{figure}

It is important to discuss the effect of screening of the Coulomb interaction between electrons by the conducting electrodes. Here, we consider the case where electron A and electron B are separated by a distance $d$ along the $x$ axis and their positions along the $z$ axis are given by $z_A$ and $z_B$, respectively. The electrode is an infinitely large plate covered with liquid helium of  depth $d_\mathrm{He}$, as shown in Fig.~\ref{fig:screening}(a).
The potential energy due to the Coulomb interaction of the two electrons is written as $U_C=\frac{e^2}{4 \pi \epsilon_0} \frac{1}{\sqrt{d^2+(z_B-z_A)^2}}$.  When the helium depth is small enough, the electrons are so close to the electrode that the interaction between the image charge induced on the electrode and the adjacent electron is non-negligible (the screening effect). In this case, the potential created by the Coulomb interaction of the two electrons is reformulated as $U_C=\frac{e^2}{4 \pi \epsilon_0} \left( \frac{1}{\sqrt{d^2+(z_B-z_A)^2}}-\frac{1}{\sqrt{d^2+(2  d_\mathrm{He}+z_B+z_A)^2}} \right) $.  With $d=0.88~\mu$m, $d_\mathrm{He}=140~$nm, the Coulomb interaction becomes $10^{-6}$ times smaller due to the screening effect, which means a much lower temperature is required for the electrons to form a Wigner crystal. 

Although the screening effect diminishes the Coulomb interaction drastically in terms of the formation of a Wigner crystal  when $d_\mathrm{He} \ll d$, it may even enhance the electric dipole-dipole interaction~\cite{Lea2000}. The leading term of the Coulomb interaction that contributes to the electric dipole-dipole interaction  without the screening effect is $\frac{e^2}{4 \pi \epsilon_0} \frac{z_A z_B}{d^3}$ and the corresponding Hamiltonian of the electric dipole-dipole interaction $W'$ is given in the main text.  The leading term of the Coulomb interaction from the screening effect is  $\frac{e^2}{4 \pi \epsilon_0} \frac{z_A z_B(d^2-8 d_\mathrm{He})}{D^5}$ with $D=\sqrt{d^2+4 d_\mathrm{He}^2}$. Thus, the enhancement factor acquired by the screening effect is $\kappa=1+\frac{d^3(d^2-8 d_\mathrm{He}^2)}{D^5}$ and is shown in  Fig.~\ref{fig:screening}(b). The electric dipole-dipole interaction is doubled when $d_\mathrm{He} \ll d$ and  its minimum value $\approx 0.8$ is reached at $d_\mathrm{He} \approx d/2$.

\section{\label{sec:two_qubit_gate_appendix} Two-qubit gate}

\subsection{Controlled-phase gate}

As discussed in the main text, we realize a two-qubit gate for the spin states by applying resonant microwaves to the Rydberg states of target qubits. This can be achieved because we can excite the Rydberg states of a target qubit (qubit B for the first MW pulse in Fig.~\ref{fig:energy_diagram_and_pulses}(b)) depending on the spin state of the target qubit thanks to the Rydberg-spin interaction and depending on the Rydberg state of an adjacent qubit (qubit A for the first MW pulse) thanks to the electric dipole-dipole interaction. However, when the Rydberg transition energy difference between different spin states of the target qubit or between different Rydberg states of the adjacent qubit is small compared to the transition rate, we cannot ignore a portion of the Rydberg state being excited even when the spin state or the Rydberg state of the adjacent qubit is in the unintended states.  Furthermore, we should also take into account the phase acquired for the target qubit in unintended states through the Rydberg-spin interaction and the electric dipole-dipole interaction. Note that making $|\Delta R_i-\Delta R_j|\gg h  f_1 $ is trivial thanks to the DC Stark shift and thus we do not have to consider off-resonant excitation of the unintended qubits.

Now we consider the phase acquired during the three MW pulse sequences as depicted in Fig.~\ref{fig:energy_diagram_and_pulses}(b) and reiterated in Fig.~\ref{fig:energy_diagram_and_pulses_supplem}. For the sake of obtaining a high two-qubit gate fidelity, we first focus on the case where $\Delta b/h=4J/h=\sqrt{15} f_\mathrm{1,Ry}$. Subsequently, we explore the phase acquired for arbitary $\Delta b$ and $J$, demonstrating that the two-qubit gate fidelity indeed reaches one of its peaks when $\Delta b/h=4J/h=\sqrt{15} f_\mathrm{1,Ry}$.

\subsubsection{Analyzing the case: $\Delta b/h=4J/h=\sqrt{15} f_\mathrm{1,Ry}$}

Here we treat the case: $\Delta b/h=4J/h=\sqrt{15} f_\mathrm{1,Ry}$. When the microwave resonant with the Rydberg transition of qubit B is applied, the Hamiltonian of qubit B is given by
\begin{equation}
H=h f_\mathrm{1,Ry}\cos(2\pi f_\mathrm{MW}) s_x^B +\frac{1}{2} \Delta R_B s_z^B -\frac{1}{2} \Delta b s_z^B \sigma^B +J s_z^A s_z^B,
\end{equation}where the spin-only terms are neglected.

$\bm{\mathrm{The \ first \ MW \ pulse:}}$ For the first MW pulse, the microwave frequency is set to $f_\mathrm{MW}=f_{0,\mathrm{Ry}B}^{n_z^A=1, \sigma^B=\downarrow}$.  The Hamiltonian on the rotating frame of the angular frequency $2\pi f_\mathrm{MW} $ becomes

\begin{equation}
H=\frac{h f_\mathrm{1,Ry}}{2} s_x^B -\frac{\Delta b}{2} s_z^B \ket{\uparrow^B} \bra{\uparrow^B}
+2J\ket{2^A} \bra{2^A} s_z^B.
\end{equation}The time duration of the first MW pulse is set so that it acts as a $\pi$ pulse for the Rydberg state when the initial state of the spin state is $\ket{\downarrow^B}$.  This rotation can be denoted as $R_x^B(\pi)$. Then, the Rydberg state of qubit B after the first pulse is $ - i\ket{2^B}$.  When the initial spin state of B is $\ket{\uparrow^B}$, there is a detuning of $-\Delta b$. In this situation, as seen in footnote~\footnote{As a simpler example, where, we consider the AC Stark shift in the case where the Hamiltonian can be written as $H=\frac{h f_\mathrm{1,Ry}}{2} s_x^B+2J s_z^B\ket{2^A}\bra{2^A}$, where the notations are the same as in the main text. We denote $\ket{n_z^B(0)}$ as the initial state of qubit B. When the initial state of qubit A is $\ket{1^A}$, the state becomes $\ket{1^A}R^B_x(h f_\mathrm{1,Ry} t_p)\ket{n_z^B(0)}$ after time $t_p$. When the initial state of qubit A is $\ket{2^A}$, the state becomes $\ket{2^A}R^B_{(\cos\upsilon,0,\sin\upsilon)}(2 \pi \sqrt{ f_\mathrm{1,Ry}^2+ \left( \frac{4J}{h} \right)^2} t_p)\ket{n_z^B(0)}$ with $\upsilon=\arctan \frac{4J}{h f_\mathrm{1,Ry}}$ after time $t_p$. Here, $R^\circ_\square(\Xi)$ is the rotation operator for the spin state of qubit $\circ=A, B$ with  angle $\Xi$ along the $\square$ axis.}, the first pulse acts as $R^B_{(\cos\upsilon_1,0,\sin\upsilon_1)}(4\pi)$ with $\upsilon_1=\arctan \frac{-\Delta b}{h f_\mathrm{1,Ry}}=\arctan(-\sqrt{15})$. Then, the Rydberg state of qubit B after the first pulse is $\ket{1^B}$ (unchanged due to the $4\pi$ rotation). 

$\bm{\mathrm{The \ second \ MW \ pulse:}}$ For the second MW pulse, the microwave frequency is set to $f_\mathrm{MW}=f_{0,\mathrm{Ry}A}^{n_z^B=2, \sigma^A=\downarrow}$.  The Hamiltonian on the rotating frame of the angular frequency $2\pi f_\mathrm{MW} $ becomes

\begin{equation}
H=\frac{h f_\mathrm{1,Ry}}{2}s_x^A -\frac{\Delta b}{2} s_z^A \ket{\uparrow^A} \bra{\uparrow^A}
-2J s_z^A\ket{1^B} \bra{1^B} .
\end{equation}The second pulse acts as a $2 \pi$ pulse when the spin state of qubit A is $\ket{\downarrow^A}$ and the Rydberg state of qubit B is $\ket{2^B}$. After this pulse, the phase is  multiplied by $-1$. When the spin state of qubit A is $\ket{\uparrow^A}$ or the Rydberg state of qubit B is $\ket{1^B}$, there is a detuning of $-\Delta b$ or $-4J$. That is, the second pulse acts as $R^A_{(\cos\upsilon_1,0,\sin \upsilon_1)}(8\pi)$. Then, the Rydberg state of qubit A after the second pulse stays the same as before the second pulse (unchanged due to the $8 \pi$ rotation). When the spin state of qubit A is  $\ket{\uparrow^A}$ and the Rydberg state of qubit B is $\ket{1^B}$ there is a detuning of $(-\Delta b-4J)/h=-2\sqrt{15} f_\mathrm{1,Ry}$. We can describe the rotation here as $R^A_{(\cos\upsilon_2,0,\sin\upsilon_2)} (\sqrt{61}\cdot 2\pi)$ with $\upsilon_2=\arctan(-2 \Delta b / h f_\mathrm{1,Ry})=\arctan(-2\sqrt{15})$. The state after the second pulse is 

\begin{equation}
    \left( \cos \left( \frac{\sqrt{61}\cdot 2\pi}{2} \right)s_0 - i \sin \left( \frac{\sqrt{61}\cdot 2\pi}{2} \right) (s_x \cos\upsilon_2+ s_z \sin \upsilon_2 ) \right) \ket{1^A}=(0.8275+0.5568i)\ket{1^A}+0.0719i\ket{2^A}, 
\end{equation}
where $s_0$ is the identity operator.    

\begin{figure}[H] 
\centering
\includegraphics[width=0.8\textwidth]{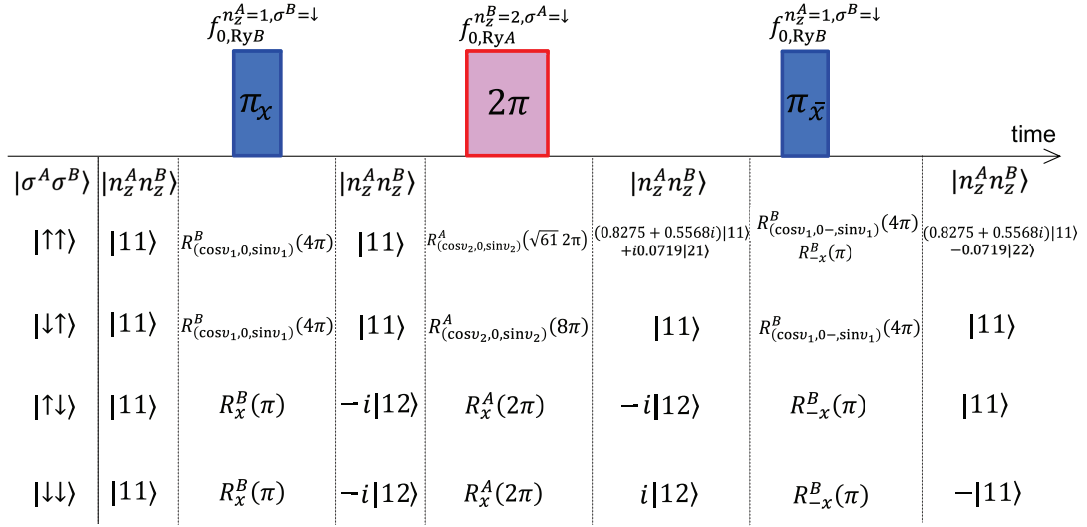}
\caption{\label{fig:energy_diagram_and_pulses_supplem} How the qubit states evolve during the three MW pulses when $\Delta b/h=4J/h=\sqrt{15}f_\mathrm{1,Ry}$.}
\end{figure}

$\bm{\mathrm{The \ third \ MW \ pulse:}}$ 
The third pulse works in the same way as the first pulse. The only difference is that the phase of the third pulse is different from the first pulse by $\pi$. 
 The Hamiltonian on the rotating frame of the angular frequency $2\pi f_\mathrm{MW} $ is

\begin{equation}
H=-\frac{h f_\mathrm{1,Ry}}{2} s_x^B -\frac{\Delta b}{2} s_z^B\ket{\uparrow^B} \bra{\uparrow^B}
+2J\ket{2^A} \bra{2^A} s_z^B.
\end{equation}
Note that when the Rydberg state of A is $\ket{2^A}$  and the spin state of B is $\ket{\uparrow^B}$, the detuning is $ -\Delta b+4J=0$. Thus, the rotation is described as $R_{-x}^B(\pi)$. In Fig.~\ref{fig:energy_diagram_and_pulses_supplem}, we summarize how the states evolve during the three pulses.

As seen in Fig.~\ref{fig:energy_diagram_and_pulses_supplem}, the total operation of a set of three pulses is described as 
\begin{equation}
\begin{pmatrix}
  0.9974 e^{ 0.1885 i \pi} & 0 & 0 & 0 \\
0 & 1 & 0 & 0 \\
0 & 0 & 1 & 0 \\
0 & 0 & 0& -1
\end{pmatrix}.
\label{eq:Cgate_leakage}
\end{equation}
Here, the Rydberg-excited states are treated as leakage states and the space is defined by the spin states of qubits A and B: $\ket{\uparrow \uparrow}$, $\ket{\downarrow \uparrow}$, $\ket{\uparrow \downarrow }$, and $\ket{\downarrow \downarrow }$. Eq.~(\ref{eq:Cgate_leakage}) is equivalent to
\begin{equation}
\begin{pmatrix}
e^{ 0.1885 i \pi}& 0 & 0 & 0 \\
0 & 1 & 0 & 0 \\
0 & 0 & 1 & 0 \\
0 & 0 & 0& -1
\end{pmatrix}\label{eq:CP01885}
\end{equation}
with a small error due to the leakage to $\ket{2^A2^B}$. Eq.~(\ref{eq:CP01885}) can be transformed into a controlled-($1.1885 \pi$) gate, $CP_{1.1885 \pi}= \ket{\uparrow^A} \bra{\uparrow^A} \otimes \sigma_0^B+ \ket{\downarrow^A} \bra{\downarrow^A} \otimes R_z^B(1.1885 \pi)$ , where $\sigma_0$ is the identity operator, by applying single-qubit z-rotations and adding some global phases. 

\subsubsection{Exploring more general cases \label{sec:cphase_general_cases}}
Here, we explore the phase acquired in more general cases. When the qubit is rotated off-resonantly with a detuning frequency $\Delta f$ the qubit state acquires a phase given by 

\begin{equation}
    \theta(h\Delta f)=\arg\left(\cos \left(\sqrt{\Delta f^2+f_\mathrm{1,Ry}^2} \pi t_p \right)+i\frac{\Delta f}{\sqrt{\Delta f^2 +f_\mathrm{1,Ry}^2}}\sin \left(\sqrt{\Delta f^2+f_\mathrm{1,Ry}^2} \pi t_p \right)\right),
\end{equation}
where \(t_p\) is the pulse length and $\arg(\circ)$ represents the phase angle of a complex number $\circ$.

$\bm{\mathrm{The \ first \ and \ third \ MW \ pulse:}}$ For states \( \ket{\sigma^A \sigma^B}=\ket{\uparrow \uparrow}, \ket{\downarrow \uparrow} \), since the detuning is $h\Delta f=-\Delta b$, the qubit acquires a phase \(\theta(-\Delta b)\) during each MW pulse . These acquired phases can be corrected by applying single-qubit rotations around the $z$-axis to the spin state of qubit B at the end. 

$\bm{\mathrm{The \ second \ MW \ pulse:}}$ For states \(\ket{\sigma^A \sigma^B}=\ket{\downarrow \uparrow}, \ket{\uparrow \downarrow}\), the qubit acquires phases, \(\theta(-4J)\) and \(\theta(-\Delta b)\), respectively. For \(\ket{\sigma^A \sigma^B}=\ket{\uparrow \uparrow}\), the qubit acquires a phase \(\theta(-(4J+\Delta b))\). The phases acquired for \(\ket{\sigma^A \sigma^B}=\ket{\downarrow \uparrow}, \ket{\uparrow \downarrow}\) can be corrected by applying single-qubit rotations around the \(z\) axis on the spin state of qubit A at the end. However, it is important to note that these corrections introduce an additional phase to the qubit state for \(\ket{\sigma^A \sigma^B}=\ket{\uparrow \uparrow}\) as well. 

To summarize the above discussion, including the phase corrections introduced by the single-qubit rotations on the spin states at the end, the total operation of a set of three pulses is described as 
\begin{equation}
\begin{pmatrix}
   e^{ i \Theta} & 0 & 0 & 0 \\
0 & 1 & 0 & 0 \\
0 & 0 & 1 & 0 \\
0 & 0 & 0& -1
\end{pmatrix},
\label{eq:Cgate_arbitary}
\end{equation}
where $\Theta=\theta(-(4J+\Delta b)) - \theta(-4J)-\theta(-\Delta b)$. In the same manner as done in the previous subsection, Eq.~\ref{eq:Cgate_arbitary} can be transformed into a controlled-($\pi+\Theta$) gate, $CP_{\pi+\Theta}$.

In addition to the acquired phases, the off-resonant rotations unintentionally populate the Rydberg states of qubits A and B. This effect causes the leakage to \(\ket{1^A 2^B}\), \(\ket{2^A 1^B}\), and \(\ket{2^A 2^B}\) at the end of the pulse sequence, resulting in the two-qubit gate infidelity. These unintentional Rydberg populations become zero when \(\sqrt{\Delta f
^2 + f_{1,\text{Ry}}^2}t_p \) is a positive even integer~\cite{Russ2018-tn,Noiri2022-ll}. Given that \( t_p = \frac{1}{2 f_{1,\text{Ry}}} \) for the first and third MW pulses, and \( t_p = \frac{1}{f_{1,\text{Ry}}} \) for the second MW pulse, the condition reduces to \( \sqrt{(\Delta f/f_{1,\text{Ry}})^2 + 1} \) being a positive even integer for the first and third MW pulses, and a positive integer for the second MW pulse. Satisfying this condition leads to a high two-qubit gate fidelity. This justifies our choice of the parameters \(\Delta b/h = 4J/h = \sqrt{15} f_{1,\text{Ry}}\), which are also experimentally feasible as introduced in the previous section and in the main text. With this set of parameters, \(\sqrt{(\Delta f/f_{1,\text{Ry}})^2 + 1}= 4\) for states \(\ket{\sigma^A \sigma^B} = \ket{\uparrow \uparrow}, \ket{\downarrow \uparrow}\) during the first and third MW pulses and for states \(\ket{\sigma^A \sigma^B} = \ket{\downarrow \uparrow}, \ket{\uparrow \downarrow}\) during the second MW pulse. Notably, the undesired Rydberg population is only observed for the state \(\ket{\sigma^A \sigma^B} = \ket{\uparrow \uparrow}\) during the second MW pulse. In Sec.~\ref{sec:two-qubit_gate_fidelity_general_cases}, we analyze the average gate fidelity for the controlled-$(\pi+\Theta)$ gate and show that it indeed peaks when \(\sqrt{\Delta f
^2 + f_{1,\text{Ry}}^2}t_p \)  is a positive even integer.

\subsection{Controlled-$\pi$ gate \label{sec:controlled-pi}}

Following Ref.~\cite{Zhang2003-yt,Bremner2002-ot}, we found that a controlled-($-\Theta$) gate, $CP_{-\Theta}$, can be constructed out of the two times of a controlled-($\pi+\Theta$) gate, $CP_{\pi+\Theta}$, and single-qubit gates, for $0<\Theta< \frac{2}{3} \pi$. Knowing that a controlled-$\pi$ gate can be realized by $CP_\pi= CP_{\pi+\Theta} CP_{-\Theta}$, $CP_\pi$ can be constructed out of the three times of $CP_{\pi+\Theta}$ and single-qubit gates. More concretely, for $\Theta=0.1885 \pi$, we found 
\begin{equation}
     CP_{1.1885 \pi} R^{B\dagger}_y(\Xi_2)R^{B\dagger}_z(\Xi_1)CP_{1.1885 \pi}
    R^{B\dagger}_x(\Xi_0) 
    CP_{1.1885 \pi} R^B_x(\Xi_0) R^B_z(\Xi_1) R^B_y(\Xi_2)=
    \begin{pmatrix}
1 & 0 & 0 & 0 \\
0 & 1 & 0 & 0 \\
0 & 0 & -i & 0 \\
0 & 0 & 0& i,
\end{pmatrix} \label{Cpi}
\end{equation}
with $\Xi _0=0.9014 \pi$, $\Xi _1=0.0943\pi$, and $\Xi _2=0.4855\pi$
 as one of the solutions. Eq.~(\ref{Cpi}) is equal to a controlled-($\pi$) gate, $CP_{\pi}= \ket{\uparrow^A} \bra{\uparrow^A} \otimes \sigma_0^B+ \ket{\downarrow^A} \bra{\downarrow^A} \otimes R_z^B( \pi)$.

\subsection{Two-qubit gate fidelity}\label{sec:two-qubit_gate_fidelity_appendix}

\subsubsection{General cases excluding the relaxation effects \label{sec:two-qubit_gate_fidelity_general_cases}}

\begin{figure}
\centering
\includegraphics[width=\textwidth]{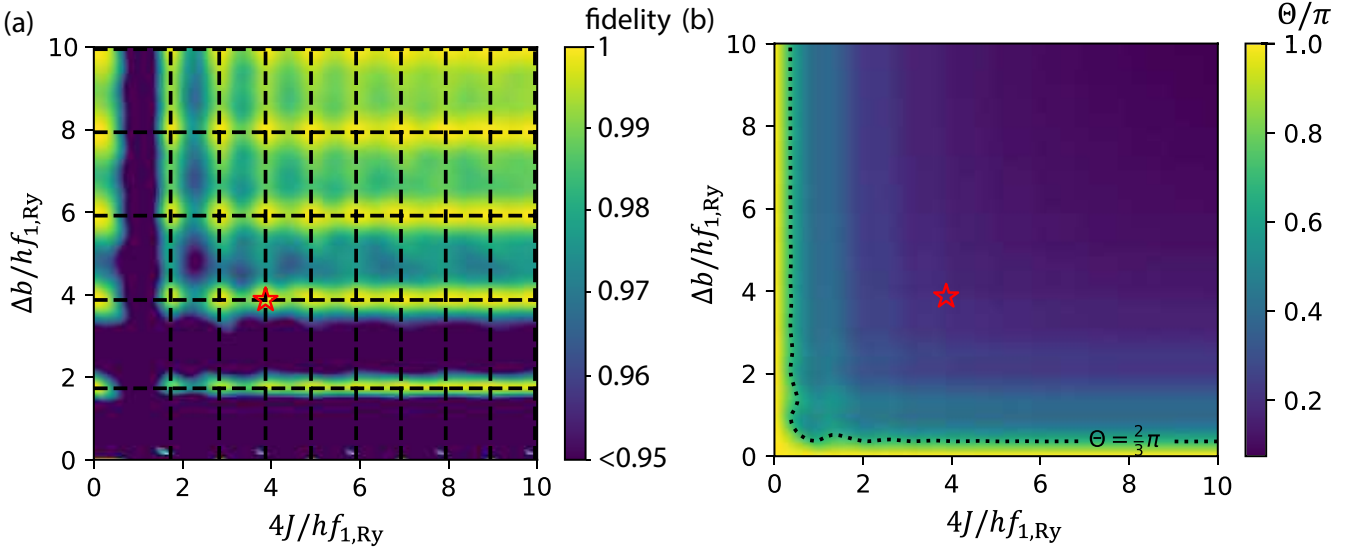}
\caption{\label{fig:Phase_Fideliy_vs_4J_Deltab}(a) Average gate fidelity of the controlled-($\pi+\Theta$) gate without taking into account the relaxation effect as a function of \(\Delta b/ h f_\mathrm{1,Ry}\) and \( 4J/ h f_\mathrm{1,Ry}\). Dashed lines represent \(\sqrt{(\Delta b/f_{1,\text{Ry}})^2 + 1}=2,4,6,8,10\) and \(\sqrt{( 4J/f_{1,\text{Ry}})^2 + 1}=2,3,4,5,6,7,8,9,10\).  (b) Phase acquired by the controlled-phase gate represented by \(\Theta/\pi\), where $\Theta$ is given by Eq.~\ref{eq:Cgate_arbitary}, as a function of \(\Delta b/ h f_\mathrm{1,Ry}\) and \( 4J/ h f_\mathrm{1,Ry}\). The dotted line shows the contour for \(\Theta=\frac{2}{3} \pi \). (a,b) The red stars mark the position where \(\Delta b/h =4J/h=\sqrt{15} f_\mathrm{1.Ry}\), which  corresponds to \(\sqrt{(\Delta b/f_{1,\text{Ry}})^2 + 1}=\sqrt{( 4J/f_{1,\text{Ry}})^2 + 1}=4\). With these parameters, the fidelity of the controlled-($\pi+\Theta$) gate is 99.92\%, with \(\Theta=0.1885 \pi\).}
\end{figure}

The average gate fidelity of the controlled-($\pi+\Theta$) gate is calculated to be 
\begin{equation}
F_2=\frac{1}{16} \left[ 4+ \frac{1}{5} \sum_{i=1}^{15} \mathrm{Tr}[\mathcal{CP}_{\pi+\Theta}^\mathrm{ideal}(\rho_i^\mathrm{spin}) \mathcal{CP}_{\pi+\Theta}^\mathrm{real}(\rho_i^\mathrm{spin})] \right].\label{eq:F_2}
\end{equation}Here $\rho_i^\mathrm{spin}$ is the Kronecker product of Pauli matrices of the spin states:
\begin{equation}
\rho_i^\mathrm{spin} = \sigma_\mu^A  \otimes \sigma_\nu^B , \;  \; \; \; \mu, \nu=0, x, y, z,
\end{equation}
 and $\sigma_0^A \otimes \sigma_0^B$  is excluded from the sum of Eq.~(\ref{eq:F_2})~\cite{Cabrera2007-aq}. $\mathcal{CP}_{\pi+\Theta}^\mathrm{ideal} (\rho_i^\mathrm{spin})$ and $\mathcal{CP}_{\pi+\Theta}^\mathrm{real} (\rho_i^\mathrm{spin})$ denote the density operators obtained after applying the controlled-($\pi+\Theta$) gate to the initial state $\rho_i^\mathrm{spin}$ in the ideal and real cases, respectively. We obtained $\mathcal{CP}_{\pi+\Theta}^\mathrm{real}  (\rho_i^\mathrm{spin})$ by numerically calculating $\mathcal{CP}_{\pi+\Theta}^\mathrm{real} (\rho_i^\mathrm{spin} \otimes  \ket{1^A}\bra{1^A} \otimes  \ket{1^B}\bra{1^B} )$ using qutip.mesolve~\cite{Johansson2013-rq}  and tracing over the Rydberg states. As detailed in Sec.~\ref{sec:cphase_general_cases}, the leakage to the states \(\ket{1^A 2^B}\), \(\ket{2^A 1^B}\), and \(\ket{2^A 2^B}\) at the end of the pulse sequence negatively impacts the gate fidelity. Note that we do not account for the effect of relaxation in this section; it is addressed in Sec.~\ref{subsec:considering_T1} for \( \Theta = 0.1885 \pi \). In Fig.~\ref{fig:Phase_Fideliy_vs_4J_Deltab} (a), we numerically calculated the gate fidelity as a function of \( \Delta b \) and \( 4J \). Notably, discernible peaks appear when leakage is minimized. This happens when \( \sqrt{(\Delta b/f_{1,\text{Ry}})^2 + 1} \) is a positive even integer and \( \sqrt{(4J/f_{1,\text{Ry}})^2 + 1} \) is a positive integer, in line with the discussions in Sec.~\ref{sec:cphase_general_cases}. Fig.~\ref{fig:Phase_Fideliy_vs_4J_Deltab} (b) shows $\Theta/ \pi$ as a function of $\Delta b/ h f_{1,\text{Ry}}$ and $ 4J/ h f_{1,\text{Ry}}$. Note that when \(\frac{2}{3} \pi < \Theta\), constructing a controlled-\(\pi\) gate cannot be achieved with only three iterations of \(CP_{\pi+\Theta}\)~\cite{Zhang2003-yt,Bremner2002-ot}.

\subsubsection{Considering the relaxation rate for the case: $\Delta b/h=4J/h=\sqrt{15}f_\mathrm{1,Ry}$ \label{subsec:considering_T1}}
In this section, we evaluate the gate fidelity for the controlled-($1.1885 \pi$) gate, taking into account the effects of the relaxation rate using the Lindblad master equation (Eq.~(\ref{Lindblad_eq})). Here, $\rho$ is the density operator of both the spin states and the two-lowest Rydberg states of qubits A and B. We take $A_1= \sqrt{T_\mathrm{1,Ry}^{-1}}\ket{1^A}\bra{2^A}$ and $A_2= \sqrt{T_\mathrm{1,Ry}^{-1}}\ket{1^B}\bra{2^B}$ as collapse operators. The Rydberg dephasing rate~\cite{Platzman1999,Dykman2003}, the spin relaxation rate, and the spin dephasing rate are much smaller than $T_\mathrm{1,Ry}^{-1}$, and thus can be ignored. The average gate fidelity of the controlled-($1.1885 \pi$) is calculated by Eq.~(\ref{eq:F_2}) taking $\Theta=0.1885 \pi$. In addition to the leakage that already appears in Eq.~(\ref{eq:Cgate_leakage}), the Rydberg relaxation degrades the fidelity. For the case considered in the main text,   $\Delta b/h=4J/h=\sqrt{15} f_\mathrm{1,Ry}= 137~$MHz (i.e., $f_\mathrm{1,Ry}=35.5$~MHz) and  $T_\mathrm{1,Ry}^{-1}=1~$MHz,  we obtained $F_2=98.5\%$. In this case, as the total rotation for the two-qubit gate is $4 \pi$ (Fig.~\ref{fig:energy_diagram_and_pulses_supplem}), the two-qubit gate time is calculated to be $2/f_\mathrm{1,Ry}=56~$ns. Fig.~\ref{fig:energy_diagram_and_pulses} illustrates the average gate fidelity of the controlled-\(1.1885 \pi\) gate with different values of \(f_\mathrm{1,Ry}\) and \(T_\mathrm{1,Ry}^{-1}\) while keeping the condition: $\Delta b/h=4J/h=\sqrt{15} f_\mathrm{1,Ry}$.

\section{\label{sec:read-out_supplem} Read-out of the qubit state}

\begin{figure}
\centering
\includegraphics[width=\textwidth]{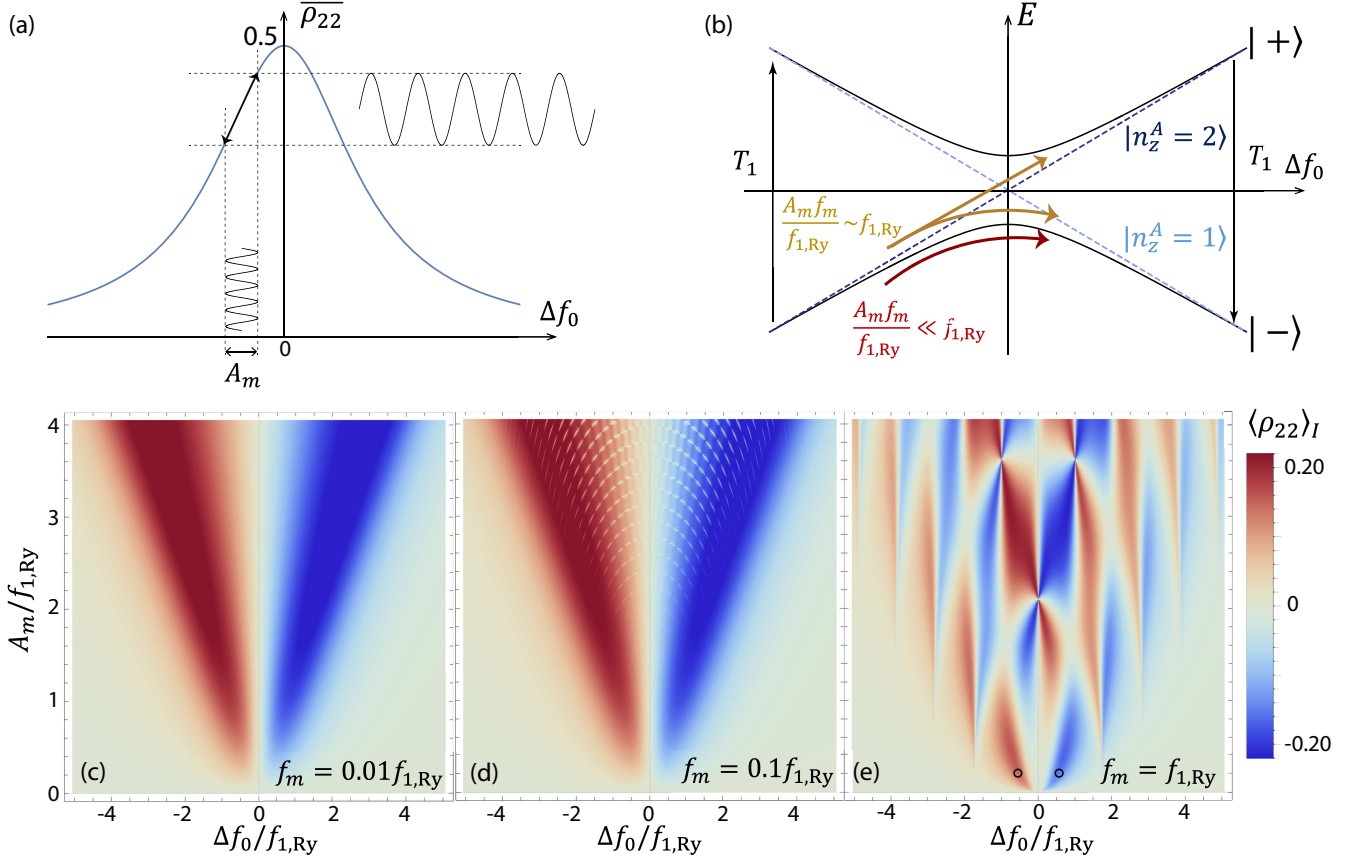}
\caption{\label{fig:anticrossing}(a)  Time-averaged probability of the electron in the excited state $\overline{\rho_{22}}$ as a function of the detuning of excitation frequency. $\overline{\rho_{22}}$ reaches its maximum value at $\Delta f_0 = 0$. Under the condition that the Rydberg transition rate is much faster than the Rydberg relaxation: $f_\mathrm{1,Ry} \gg T_\mathrm{1,Ry}^{-1}$, $T_\mathrm{2,Ry}^{-1}$, the maximum value is $\sim 0.5$. (b) Energy levels $E$ versus the MW frequency detuning $\Delta f_0$. The detuning is modulated with an amplitude of $2\pi A_m$ and a frequency of $f_m$. $\ket{\pm}$ are the eigenstates of $H_\mathrm{R}$ (Eq.~(\ref{Hamiltonian_for_detection_R})) without voltage modulation ($A_m=0$) and $\ket{n_z^A=1}$ and $\ket{n_z^A=2}$ are those without MW excitation ($f_\mathrm{1,Ry}=0$). The state-relaxation happens from $\ket{n_z^A=2}$ and $\ket{n_z^A=1}$. When $A_m f_m/f_\mathrm{1,Ry}\gg f_\mathrm{1,Ry}$, the LZ transition reaches the fast-passage limit and the electron always stays in  $\ket{n_z^A=0}$ (that is, $\rho_{22}$ always stays 0) after time $T_\mathrm{1,Ry}$.  When $A_m f_m/f_\mathrm{1,Ry}\approx f_\mathrm{1,Ry}$, the state is changed from $\ket{+}$ to $\ket{-}$ or vice versa with some probability $p$ and stays in the same state with the probability of $1-p$, which results in LZ interference patterns shown in (d,e). When $A_m f_m/f_\mathrm{1,Ry}\ll f_\mathrm{1,Ry}$, the LZ transition reaches the slow-passage limit and the state always stays in $\ket{-}$. (c-e) Numerically calculated in-phase component of the fraction in the excited state $\left<\rho_{22} \right>_I$ as a function of normalized detuning $\Delta f_0/f_\mathrm{1,Ry}$ and the modulation amplitude $A_m/f_\mathrm{1,Ry}$ for the modulation frequency $f_m=0.01 f_\mathrm{1,Ry}$,  $f_m=0.1 f_\mathrm{1,Ry}$, and  $f_m= f_\mathrm{1,Ry}$ in (c), (d), and (e), respectively. The relation between the Rydberg relaxation rate and the Rydberg transition rate is set as $T_\mathrm{1,Ry}^{-1}=2 T_\mathrm{2,Ry}^{-1}=f_\mathrm{1,Ry}/50$.  }
\end{figure}

As discussed in Sec.~\ref{sec:detection}, the readout of the spin state can be realized by spin-selectively exciting the Rydberg state and detecting the image-charge change induced by the Rydberg excitation. Here, we consider the case where we excite the electron from the Rydberg-ground state to the Rydberg-1st-excited state only when the electron is in the spin-up state. The excitation to the Rydberg-2nd-excited state or higher can be ignored since it is detuned further away $>10~\text{GHz}$. When the spin state is down, we have off-resonant Rydberg excitation with the detuning $\Delta f_0=\Delta b$ and consequently, $\Delta f_0/f_\mathrm{1,Ry}=\sqrt{15}$ assuming that we use the same MW power as for the two-qubit gate. 
As discussed below, to have the optimal measurement condition, $A_m / f_\mathrm{1,Ry}$ is set to $0.1$. As seen in Fig.~\ref{fig:anticrossing} (c-e)), the signal is observed to be nearly zero at this point and thus this off-resonant excitation can also be ignored for the readout scheme.

 Therefore, by focusing on the two lowest Rydberg states and the spin-up state, we can write the Hamiltonian of qubit A under MW excitation with the voltage modulation as 
\begin{equation}
    H'=h f_\mathrm{0,Ry}s_z^A+eE^\mathrm{MW} z \cos(2\pi f_\mathrm{MW}t)+eE^m z\cos(2 \pi f_m t) \label{Hamiltonian_for_detection}.
\end{equation}

In the rotating reference frame of the frequency $f_\mathrm{MW}$, $H'$ (Eq.~(\ref{Hamiltonian_for_detection})) can be approximated by 
\begin{equation}
H_\mathrm{R}= h \Delta f(t) s_z^A + hf_\mathrm{1,Ry}s_x^A, \label{Hamiltonian_for_detection_R}
\end{equation}
where  $A_m=eE^m (z_{22}-z_{11})/h$. The time-dependent eigenstates of $H_\mathrm{R}$ (Eq.~(\ref{Hamiltonian_for_detection_R})) can be expressed with a density operator $\rho(t)=\sum_{(\eta,\zeta)=(1,2)} \rho_{\eta \zeta}(t) \ket{n_z^i=\eta}\bra{n_z^i=\zeta}$. The solution to the equation of motion with the Hamiltonian  $H_\mathrm{R}$ (Eq.~(\ref{Hamiltonian_for_detection_R})) is known as exhibiting a sequence of Landau-Zener transitions~\cite{Landau1932,Zener1932,Shevchenko2010}. 
The density operator can further be rewritten as $\rho(t) =\frac{1}{2} s_0 + \frac{1}{2} \vec{s} \cdot\vec{r} $  where $\vec{s}=(s_x,s_y,s_z)$ and $\vec{r}=(r_x,r_y,r_z)$. The time evolution of this density operator under $H_\mathrm{R}$ (Eq.~(\ref{Hamiltonian_for_detection_R})) can be described by the Bloch equations as
\begin{align*} 
\frac{d r_x}{dt} &=  -\frac{1}{T_\mathrm{2,Ry}} r_x- 2\pi \Delta f(t) r_y \\ 
\frac{d r_y}{dt} &=  -\frac{1}{T_\mathrm{2,Ry}} r_y+ 2 \pi \Delta f(t) r_x - 2 \pi f_\mathrm{f_1,Ry} r_z \\ 
\frac{d r_z}{dt} &=  -\frac{1}{T_\mathrm{1,Ry}} (r_z+1)+ 2 \pi f_\mathrm{f_1,Ry} r_y,
\end{align*}where $\Delta f(t)$ is defined in Sec.~\ref{sec:detection}. We numerically solved the above differential equations and calculated $\rho_{22}(t)=\frac{r_z(t)+1}{2}$. From this, we obtain the in-phase component of $\rho_{22}$, which is plotted in Fig.~\ref{fig:anticrossing} (c--e).

As discussed in Sec.\ref{sec:detection}, the read-out signal is characterized by $\Delta C = \alpha' \Delta q\frac{\left<\rho_{22} \right>I}{A_m}$. There is an optimal set of parameters for $\frac{\left<\rho_{22} \right>I}{A_m}$ to be maximized. As illustrated in Fig.\ref{fig:anticrossing} (a, c-e), the signal is zero when exactly on resonance but reaches its maximum when positioned on the shoulder of the resonance peak. If the modulation frequency $f_m$ is significantly greater than the Rabi frequency of the Rydberg state $f_\mathrm{1,Ry}$, the electron consistently remains in its initial state, as depicted in Fig.\ref{fig:anticrossing} (b). Consequently, there is no change in the image charge. We found that the optimal point occurs when $f_m  \approx f_\mathrm{1,Ry}$, the MW frequency is slightly detuned, and the modulation amplitude $A_m$ is considerably small, which corresponds to the black circle positions in Fig.\ref{fig:anticrossing} (e), where $\left<\rho_{22} \right>_I \approx 0.1$ and $A_m = f_\mathrm{1,Ry}/10=5$MHz, which gives $\Delta C \approx 60$~aF.

\bibliography{library.bib}

\end{document}